%% file: main.tex
\documentclass[11pt,a4paper]{article}
\usepackage{jheppub}

\usepackage{graphicx}
\usepackage{amsmath}
\usepackage{caption}
\usepackage[labelformat=empty]{subcaption}
\usepackage{epstopdf}
\usepackage[separate-uncertainty=true]{siunitx}
\usepackage{adjustbox}
\usepackage{tikz}

\input{sections/makros.tex}

\title{Mixed QCD$\otimes$QED corrections to on-shell $Z$ boson production at the LHC
  }

\author[a]{Maximilian~Delto,}
\author[a]{Matthieu~Jaquier,}
\author[a]{Kirill~Melnikov,}
\author[b]{Raoul~R\"ontsch}

\affiliation[a]{Institute for Theoretical Particle Physics, KIT, Karlsruhe, Germany}
\affiliation[b]{Theoretical Physics Department, CERN, 1211 Geneva 23, Switzerland}

\emailAdd{maximilian.delto@kit.edu}
\emailAdd{matthieu.jaquier@kit.edu}
\emailAdd{kirill.melnikov@kit.edu}
\emailAdd{raoul.rontsch@cern.ch}

\preprint{TTP19-029,\;P3H-19-032,\;CERN-TH-2019-148} 

\abstract{We compute the mixed QCD$\otimes$QED corrections to the production of on-shell $Z$ bosons at the LHC at a fully-exclusive level.
  We also include the factorised NLO QCD correction to $Z$ boson production and
  NLO QED correction to $Z$ boson decay into two leptons.
  We make use of an abelianised version of the nested soft-collinear subtraction formalism to  perform this computation. 
  We study the phenomenological impact of 
  the mixed QCD$\otimes$QED corrections for a number of observables relevant for LHC phenomenology.}

\begin{document}
\maketitle
\flushbottom

\input{sections/01_introduction.tex}

\input{sections/02_technicalities.tex}

\input{sections/03_results.tex}
\input{sections/04_conclusion.tex}

\end{document}

%% file: sections/makros.tex
\newcommand\fverb{\setbox\fverbbox=\hbox\bgroup\verb}
\newcommand\fverbdo{\egroup\medskip\noindent%
			\fbox{\unhbox\fverbbox}\ }
\newcommand\fverbit{\egroup\item[\fbox{\unhbox\fverbbox}]}
\newbox\fverbbox

\newcommand{\ba}{\begin{equation}}
\newcommand{\ea}{\end{equation}}
\newcommand{\be}{\begin{eqnarray}}
\newcommand{\ee}{\end{eqnarray}}

\def\Tr{\, {\rm Tr}}

\newcommand{\gev}{\mathrm{GeV}}

\newcommand{\as}{\alpha_S}

\newcommand{\DIFFL}{\mathop{}\!\mathrm{d}}

%% file: sections/01_introduction.tex
\section{Introduction}
The recent years have seen a transition of the  Large Hadron Collider (LHC)
from a discovery machine to a precision machine. The reason for this is the  absence of a direct
observation of New Physics which, even after the discovery of the Higgs boson \cite{Aad:2012tfa,Chatrchyan:2012xdj}, 
is needed to clarify questions left open by the Standard Model.
Forthcoming searches for physics Beyond the Standard Model (BSM) will focus 
on systematic  studies of possible small deviations from Standard Model predictions in precision observables. 
A reliable theoretical description of such observables within the Standard Model is  an important prerequisite for
the success of this research program. 

A case in point  is the  hadronic production of charged leptons via a virtual photon and/or the $Z$ boson,
the celebrated Drell-Yan (DY) process \cite{Drell:1970wh} (see \cite{Mangano:2015ejw} for a review).
Its high production rate and distinct signature make it extremely  useful for luminosity monitoring \cite{Dittmar:1997md, Khoze:2000db, Giele:2001ms} and detector calibration \cite{Haywood:1999qg}. Being theoretically well-understood, this process is also suited for electroweak (EW) precision physics, such as the measurement of the weak mixing angle \cite{Haywood:1999qg, Sirunyan:2018swq}.
Moreover, DY production is used in parton distribution function (PDF) fits~\cite{Harland-Lang:2014zoa,Ball:2017nwa,Alekhin:2016uxn,Hou:2019qau}
and  for searches for New Physics at high energies~\cite{Farina:2016rws}.
In such analyses, the rapidity distribution of the $Z$ boson and the dilepton invariant mass, respectively, are of particular interest.

The inclusive next-to-leading order (NLO) QCD corrections to Drell-Yan production were first computed four decades ago \cite{Altarelli:1979ub}.
Inclusive results  at next-to-next-to-leading order (NNLO) in QCD have also been known for many years~\cite{Hamberg:1990np, vanNeerven:1991gh, Harlander:2002wh}.
Arbitrary infrared safe
kinematic distributions are also available  through  NNLO QCD accuracy \cite{Catani:2009sm, Melnikov:2006di, Melnikov:2006kv, Gavin:2010az, Gavin:2012sy, Boughezal:2016wmq}.
In addition, threshold effects at next-to-next-to-next-to leading order (N$^3$LO) have been studied in Refs.~\cite{Ahmed:2014cla, Catani:2014uta}.
EW corrections to $pp\rightarrow\ell^+\ell^-$ were computed in Refs.~\cite{Baur:2001ze, Baur:1997wa}.

Recently, an important milestone in the quest for high precision theoretical  predictions for LHC processes
has been reached with the calculation of Higgs boson production in hadronic collisions at N$^3$LO QCD
\cite{Mistlberger:2018etf}. Since techniques developed in the course of that calculation put
the N$^3$LO QCD corrections to the Drell-Yan process within reach, it becomes important to
know the mixed QCD-EW $\mathcal{O}(\alpha_s\alpha)$ corrections as well since, based on the
sizes of strong and EW coupling constants, one expects both contributions to be comparable in
magnitude. The computation of mixed QCD-EW $\mathcal{O}(\alpha_s\alpha)$ corrections requires the evaluation of complicated
two-loop diagrams with up to two  massive propagators, as well as the respective real-virtual and double-real contributions where a photon and/or a parton is emitted in the final state. All these contributions contain intertwined QCD and QED singularities, which need to be extracted and cancelled properly.

Several ingredients required for the calculation of $\mathcal{O}(\alpha_s\alpha)$ corrections to DY production
have already appeared
in the literature. In Ref.~\cite{Bonciani:2016wya} integrated double-real contributions to
the production of a single on-shell gauge boson have been computed using the method of reverse unitarity \cite{Anastasiou:2002yz}.
Furthermore, the two-loop master integrals needed for the double-virtual contributions
were recently presented in Ref.~\cite{Heller:2019gkq}. Nevertheless, up to now, the
various ingredients have not been combined  in a way that allows one to compute physical observables. 

Given the absence of the full calculation, different approximations have been used in
the past to estimate mixed QCD$\otimes$EW corrections.
In Ref.~\cite{Li:2012wna} NNLO QCD corrections
have been combined additively with the NLO EW ones.
Results for genuine mixed QCD$\otimes$EW effects in the leading-logarithmic approximation 
were presented in Ref.~\cite{Barze:2013fru},
under the assumption that the NLO QCD and EW  corrections factorise.
This work also included the matching of the NLO QCD and EW corrections to
QCD parton showers and  multiple photon emissions.

Although the generic Drell-Yan process $pp \to l^+l^-$ is the target of many experimental analyses,
the theoretical
description of the {\it on-shell} production of $Z$ bosons
$pp \to Z \to l^+l^-$ offers significant simplifications.
Indeed, for an on-shell $Z$ boson, virtual and real contributions
that connect incoming partons and outgoing leptons are suppressed by
the ratio of the $Z$ boson  width to its mass,  $\Gamma_Z/M_Z$.
The EW corrections to the production of a lepton pair via an on-shell $Z$ boson
can thus be separated into gauge-invariant subsets according to whether the correction
is associated with the production of the $Z$ boson  (initial) or with its  decay (final).
Similarly,  mixed QCD$\otimes$EW corrections can be
divided into an initial-initial and an initial-final contribution.
Based on the magnitude
of various contributions observed  at next-to-leading order, the initial-final corrections were 
argued to provide the dominant contribution to mixed QCD$\otimes$EW
corrections \cite{Dittmaier:2014qza} and  were subsequently studied in Ref.~\cite{Dittmaier:2015rxo}.

Recently,  mixed QCD$\otimes$QED corrections to the inclusive
production of an on-shell $Z$ boson in hadronic collisions have been computed  in Ref.~\cite{deFlorian:2018wcj}.
These corrections provide a gauge-invariant subset of the initial-initial QCD$\otimes$EW corrections; they can be obtained
from the known NNLO QCD corrections to on-shell $Z$ boson production through an abelianisation procedure. 
The mixed QCD$\otimes$QED corrections computed in  Ref.~\cite{deFlorian:2018wcj}
turned out to be quite significant at the LHC, being smaller than the NNLO QCD corrections by only a  factor of three. 
This rather  modest suppression of the initial-initial QCD$\otimes$QED corrections to the inclusive cross section 
relative  to NNLO QCD corrections  makes it interesting to study the mixed corrections 
 to more  exclusive observables. 

 In addition, the computation of mixed QCD$\otimes$QED corrections to $Z$ boson
 production is an important
 step towards the calculation  of the such corrections to the production of an on-shell $W$ boson at the LHC,
 which is of high relevance  for the $W$-mass determination. Indeed, while interactions of $W$ bosons with photons introduce additional subtleties in the computation of such  corrections  compared to  the $Z$ boson  case, understanding
 the infrared structure of mixed corrections in $pp \to Z$ is a prerequisite for the analysis
 of mixed corrections to $pp \to W^\pm$.

 The goal of this paper is to present the calculation of the fully-differential mixed QCD$\otimes$QED corrections to the production of an on-shell $Z$ boson in hadronic collisions
 (the \textit{initial-initial} corrections).
 This contribution features the most complex structure of
 infrared singularities and, for this reason, represents an important step towards the computation of full QCD$\otimes$EW corrections to $Z$ boson production.
 In addition to  mixed initial-initial corrections, we also compute 
 {\it initial-final} corrections  that arise through an interplay of  QCD corrections to $Z$ production and QED corrections
 to its decay.
Comparisons of the two contributions for various observables
 will allow us to quantify the degree of dominance of the initial-final corrections over the initial-initial ones.

 The calculation is performed by extending the nested soft-collinear subtraction scheme presented in Ref.~\cite{Caola:2017dug}
 for NNLO QCD computations through the  abelianisation procedure of  Ref.~\cite{deFlorian:2018wcj}. We make use of the
 NNPDF3.1luxQED PDF set \cite{Manohar:2016nzj,Manohar:2017eqh,Bertone:2017bme}, whose evolution is correct through $\mathcal{O}(\alpha_s\alpha)$; this enables us to remove collinear singularities from initial state radiation in a consistent way. We use the resulting code to study the impact of the QCD$\otimes$QED corrections on several distributions of phenomenological interest, including
 the transverse momentum and the rapidity spectra of the $Z$ boson, the transverse momentum distributions  of
 leptons and distributions in one of the so-called Collins-Soper angles $\theta^*$~\cite{Collins:1977iv}.

 The rest of this paper is organised as follows. In Section \ref{technicalities} we briefly summarise
 technical details  of the calculation and explain how a transition
 from NNLO QCD to mixed QCD$\otimes$QED corrections is accomplished.
 In Section \ref{results} we present phenomenological results. We conclude in Section \ref{conclusion}.

%% file: sections/02_technicalities.tex
\section{Technical aspects of the calculation}\label{technicalities}

Our goal is to compute mixed QCD$\otimes$QED corrections starting from the existing implementation
of NNLO QCD corrections to  the on-shell $Z$ boson production at a fully-differential level \cite{Caola:2019nzf}. We describe
the  relevant technical aspects of this calculation  in this section.

\subsection{Preliminary remarks}
As we mentioned in the introduction,
the mixed QCD$\otimes$QED corrections to the production cross section of an on-shell $Z$ boson
and its subsequent leptonic decay 
 can be divided into  \textit{initial-initial} and  \textit{initial-final} contributions,
while the interference between production and decay sub-processes is suppressed by the ratio of the $Z$ boson
width to its mass, $\Gamma_Z/M_Z$.
The required  initial-final matrix elements can be constructed
from the $\mathcal{O}(\alpha_s)$  and $\mathcal{O}(\alpha)$ helicity amplitudes for the production  and decay sub-processes, respectively.
The infrared singularities arising from these corrections can be handled using standard NLO techniques;
we employ the Frixione-Kunszt-Signer subtraction scheme~\cite{Frixione:1995ms,Frixione:1997np} to deal with these.
The only subtlety is that  spin correlations between production and decay processes caused by 
the spin-one nature of the intermediate $Z$ boson need to be properly accounted for.

The initial-initial corrections pose a greater challenge.
The main ingredients required for the computation of these corrections are:
\begin{itemize}
\item the tree-level matrix elements for the {\it parton-initiated} processes $q\bar{q}\rightarrow Z+\gamma+g$, $q\bar{q}\rightarrow Z+q+\bar{q}$, $qq\rightarrow Z+q+q$ and $qg\rightarrow Z+q+\gamma$;
\item the tree-level matrix elements for the {\it photon-initiated} processes $q\gamma\rightarrow Z+q+g$ and $g\gamma\rightarrow Z+q+\bar{q}$;
\item the matrix elements for the one-loop  QED correction to the {\it parton-initiated} processes $q\bar{q}\rightarrow Z+g$ and $qg\rightarrow Z+q$;
\item the matrix elements for the one-loop  QCD correction to the {\it parton-initiated} process $q\bar{q}\rightarrow Z+\gamma$ ;
\item the matrix elements for the one-loop  QCD correction to the {\it photon-initiated} process $q\gamma\rightarrow Z+q$;
\item the matrix elements for the two-loop mixed QCD$\otimes$QED correction to $q\bar{q}\rightarrow Z$ production.
\end{itemize}
All these matrix elements contain infrared singularities due to soft and/or collinear emissions of gluons, photons and quark-antiquark pairs.
These singularities have to be regularised and removed in an appropriate subtraction scheme,
yielding a fully-differential description of on-shell $Z$ boson production suitable for numerical integration.
We achieve this goal by abelianising the NNLO QCD calculation of $Z$ boson production
performed within the  nested soft-collinear subtraction scheme~\cite{Caola:2017dug,Caola:2018pxp,Delto:2019asp,Caola:2019nzf,Caola:2019pfz}.
The abelianisation procedure was recently described in Ref.~\cite{deFlorian:2018wcj};
in essence this is a set of
rules that allows one to replace the $SU(3)$ colour factors in  the  NNLO QCD formulas in such a way that
the computation of  mixed QCD$\otimes$QED corrections becomes possible.
We discuss these replacement rules in the next section. 

\subsection{The mapping of colour factors}
There are three $SU(3)$ colour factors, $C_F^2$, $C_AC_F$ and $C_FT_R$,  which appear in NNLO QCD corrections to $Z$ boson production.
For the purpose
of turning a NNLO QCD computation into a computation of QED$\otimes$QCD corrections, these colour
factors require different modifications. We discuss them in turn. 

\begin{figure}[h]
 \centering
 \begin{subfigure}[l]{0.35\textwidth} \includegraphics[width=\textwidth]{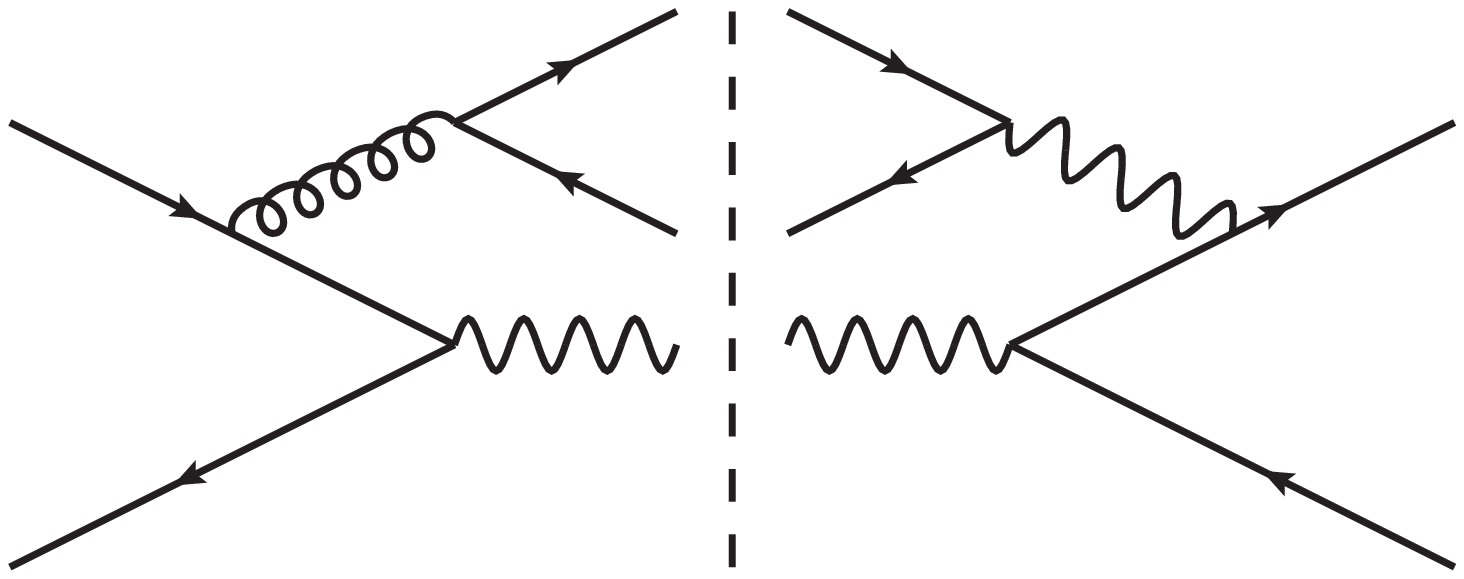} \subcaption{(a)} \end{subfigure}$\qquad\qquad\qquad$
 \begin{subfigure}[l]{0.35\textwidth} \includegraphics[width=\textwidth]{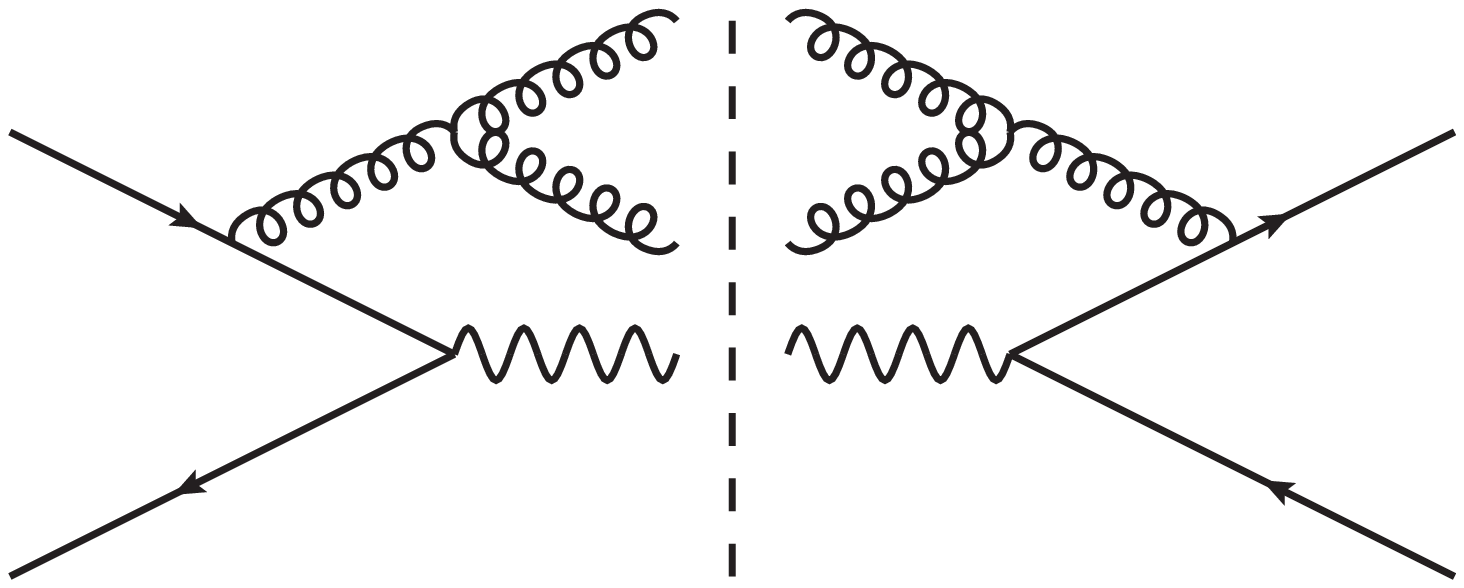} \subcaption{(b)} \end{subfigure}
 \caption{Contributions to the colour factors $C_FT_R$ (left)  and $C_AC_F$ (right).}\label{remain}
\end{figure}

We first consider the NNLO QCD computation of  $Z$ boson production
from either a quark-antiquark or a quark-quark initial state. 
The colour factor $C_FT_R$ appears 
in diagrams with two disjoined quark lines, see e.g. Fig.~\ref{remain}(a). When such
diagrams are squared and sums  over colours of initial- and final-state
particles are computed, two independent colour traces appear.
These  colour traces are of the form $\Tr(T^aT^b)\Tr(T^bT^a)=C_FT_R$. When a gluon is replaced by a photon in these diagrams 
the colour traces become  $\Tr(T^a) \Tr(T^a) $ and vanish. Similarly, 
partonic processes $q_i q_j \to Z + q_i+ q_j$ for $i \ne j$  that contribute at  NNLO QCD  become  irrelevant for ${\cal O}(\alpha \alpha_s)$
corrections.  We note that the consequence of that is  the absence of terms  
  proportional to products of two different electric quark charges in mixed QCD$\otimes$QED contributions.
  We conclude that
  NNLO QCD contributions proportional  to the colour factor $C_FT_R$ have no counter-parts in  the computation of
  QED$\otimes$QCD corrections and need to be removed. We achieve this by setting $T_R$ to zero in the
  expressions for NNLO QCD corrections provided in Ref.~\cite{Caola:2019nzf}.

The case  of  the colour factor $C_A C_F$ is very similar. The colour factors $C_A$ originate either from diagrams with three-gluon
vertices (see Fig.~\ref{remain}(b))   or from the non-commutative nature of
generators of $SU(3)$ colour algebra in the fundamental representation.  Neither  of these
issues  apply to the case of mixed QCD$\otimes$QED corrections. The corresponding contributions can be eliminated by setting $C_A$ to zero
in the  NNLO QCD computation. 

Finally, we need to understand how $C_F^2$ colour factors should
be modified for the purpose of computing mixed QCD$\otimes$QED corrections.
We consider a collision of a quark $q$ with an anti-quark $\bar q$, assume that the electric charge of the quark is $e_q$,
and discuss a few illustrative examples.

\begin{figure}[h]
 \centering
 \begin{tikzpicture}[scale=1.0]
  \node at (-5.0,2.4) {\includegraphics[width=0.26\textwidth]{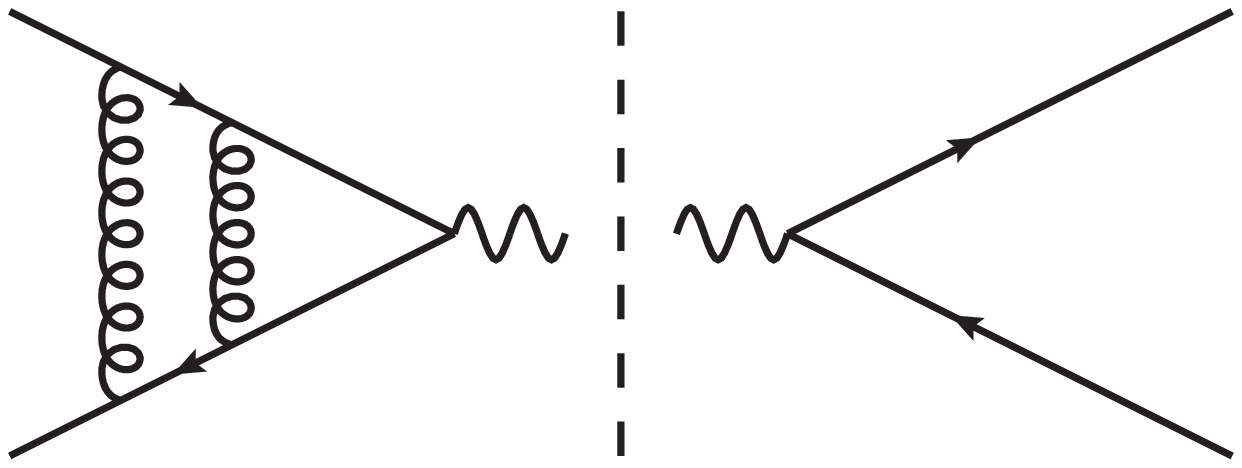}};
  \node at (1.0,2.4) {\includegraphics[width=0.26\textwidth]{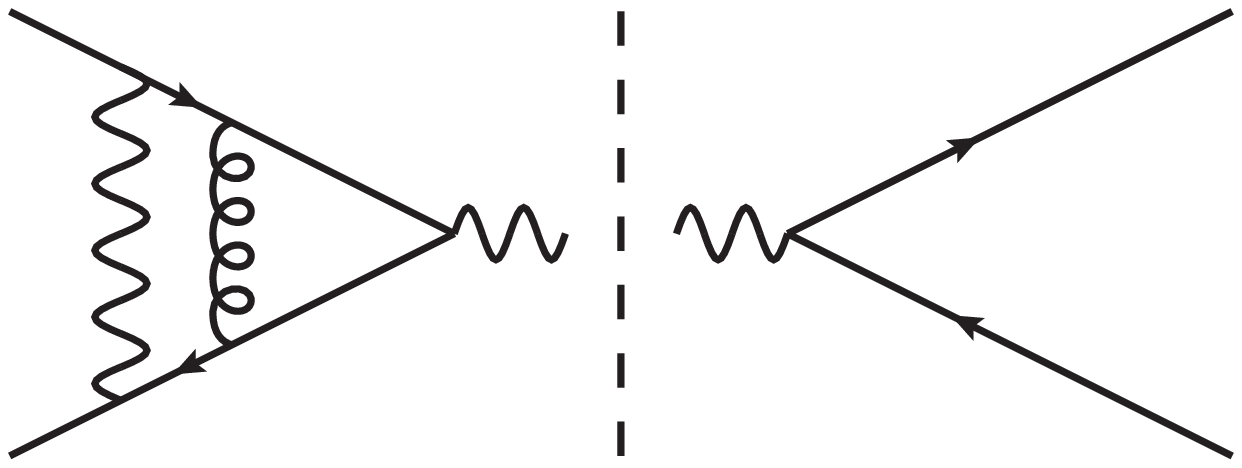}};
  \node at (6.0,2.4) {\includegraphics[width=0.26\textwidth]{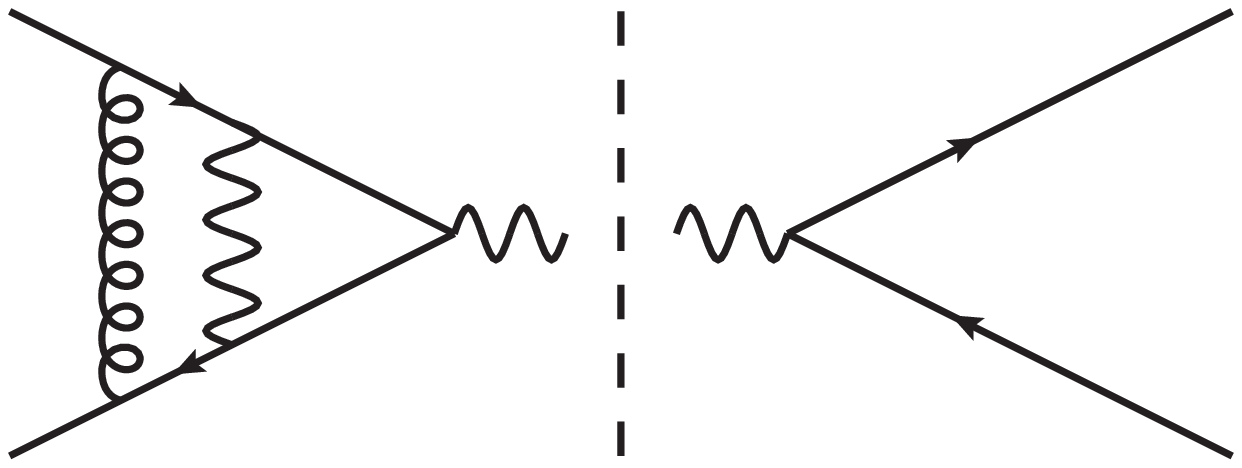}};
  \node at (3.2,2.2) {,};
  \draw[->,black,thick] (-2.5,2.4) -- (-1.5,2.4);
  \node at (-5.0,0.0) {\includegraphics[width=0.26\textwidth]{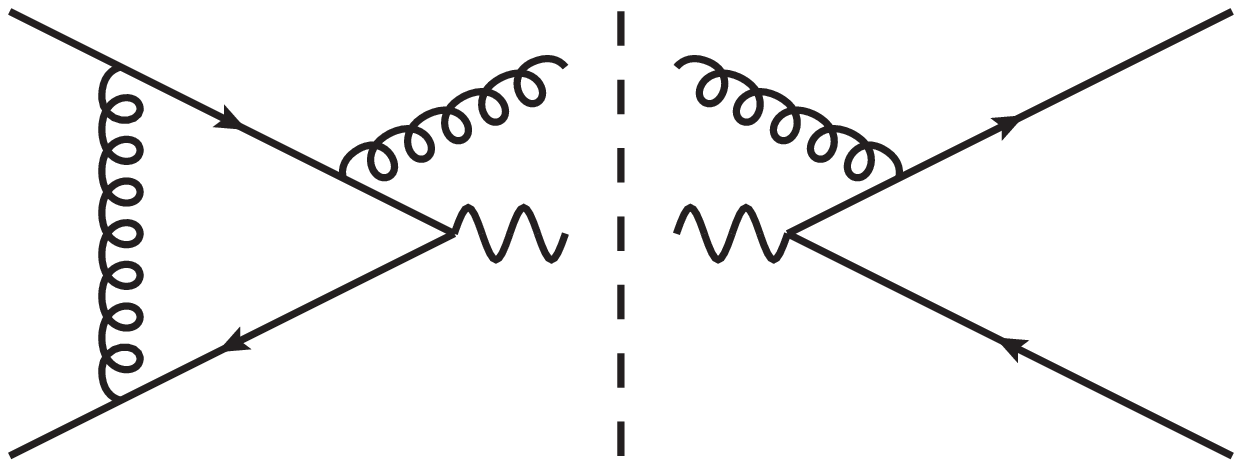}};
  \node at (1.0,0.0) {\includegraphics[width=0.26\textwidth]{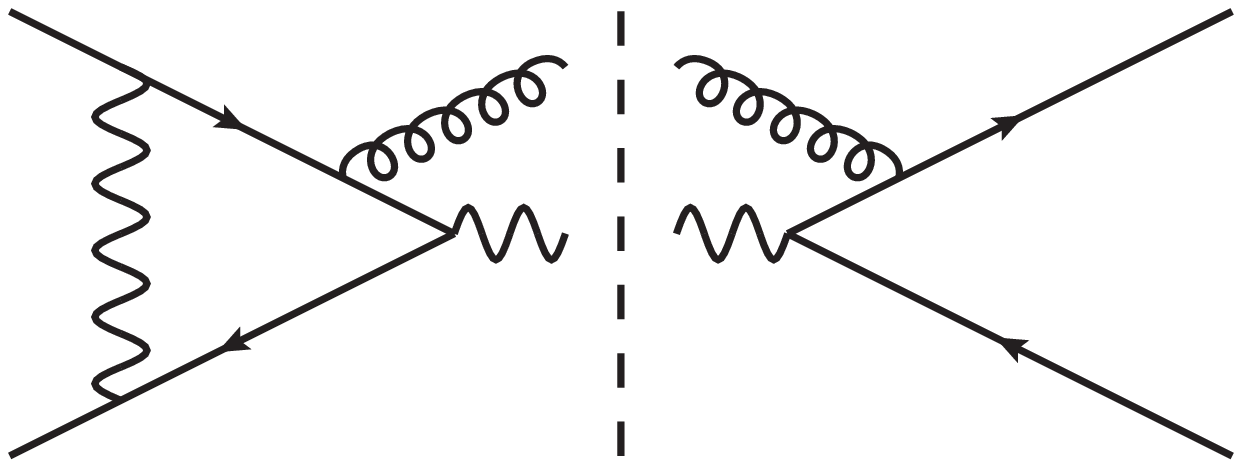}};
  \node at (6.0,0.0) {\includegraphics[width=0.26\textwidth]{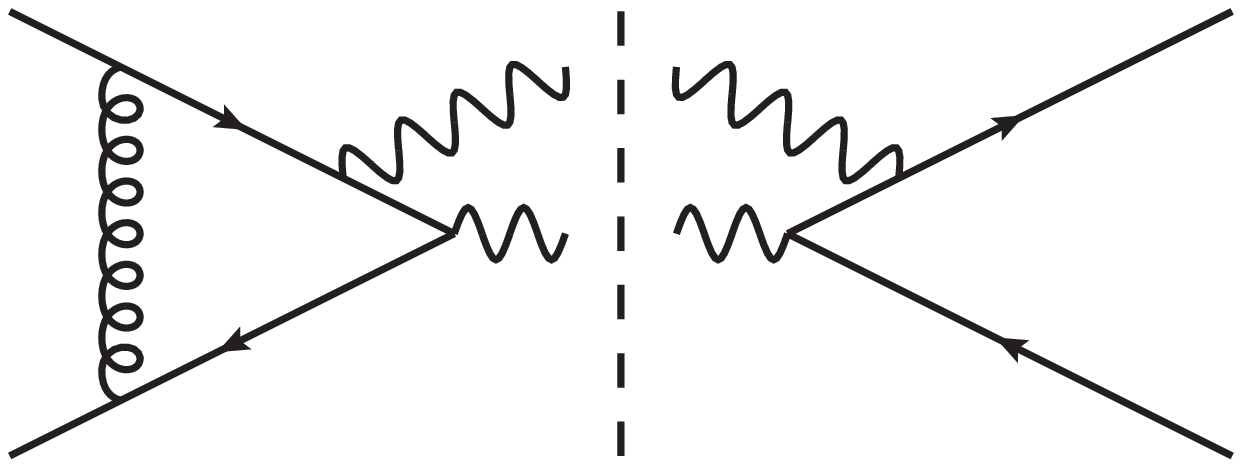}};
  \node at (3.2,-0.2) {,};
  \draw[->,black,thick] (-2.5,0.0) -- (-1.5,0.0);
  \node at (-5.0,-2.4) {\includegraphics[width=0.26\textwidth]{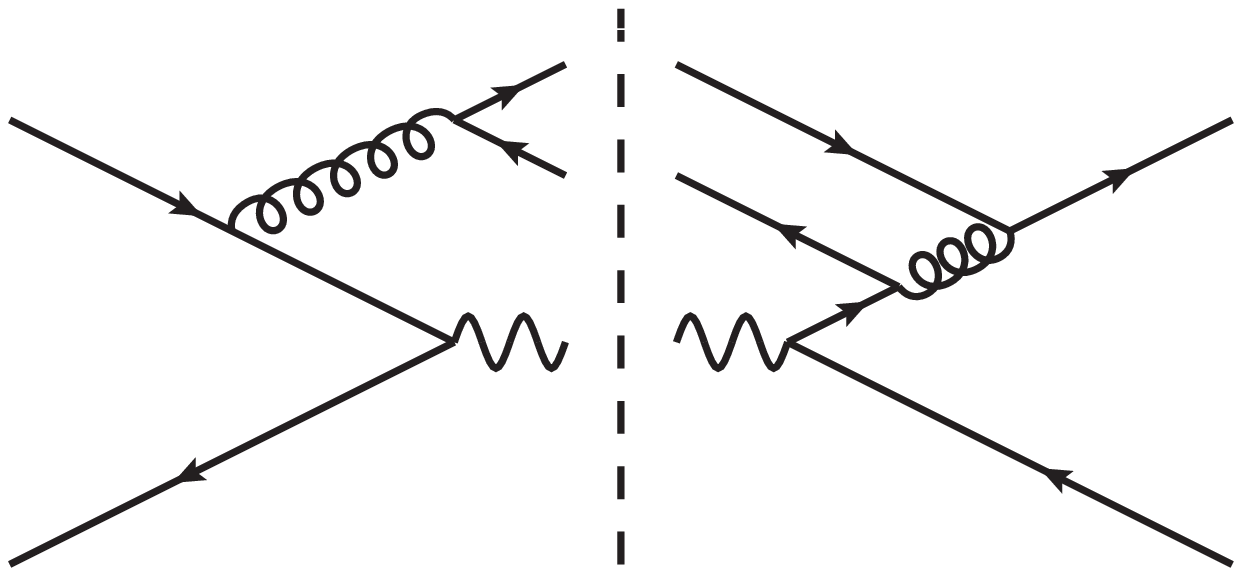}};
  \node at (1.0,-2.4) {\includegraphics[width=0.26\textwidth]{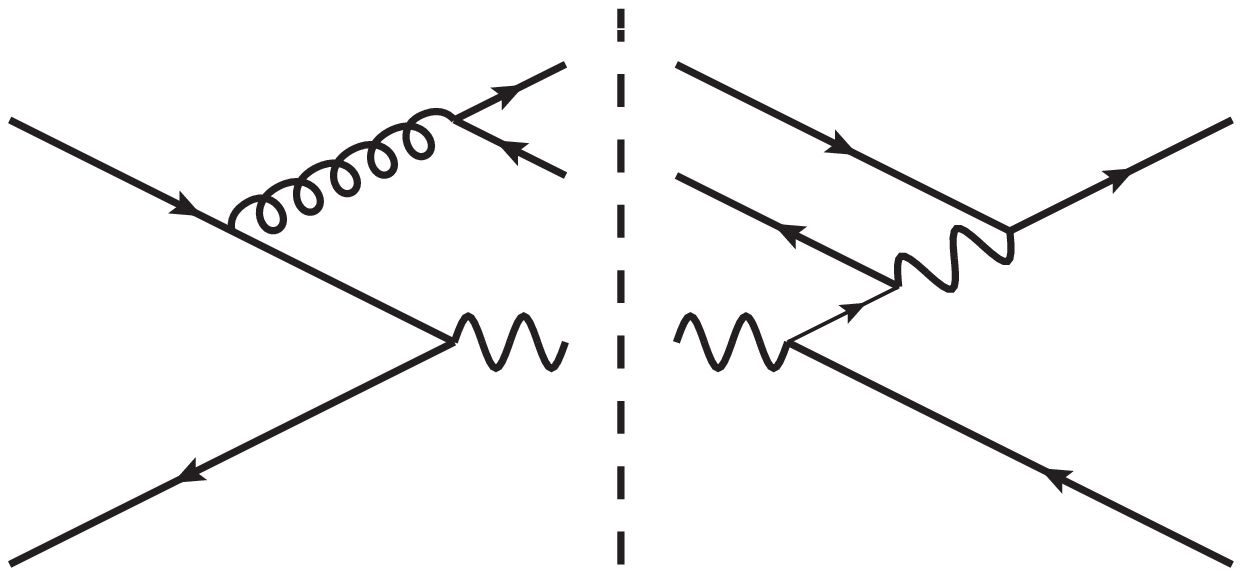}};
  \node at (6.0,-2.4) {\includegraphics[width=0.26\textwidth]{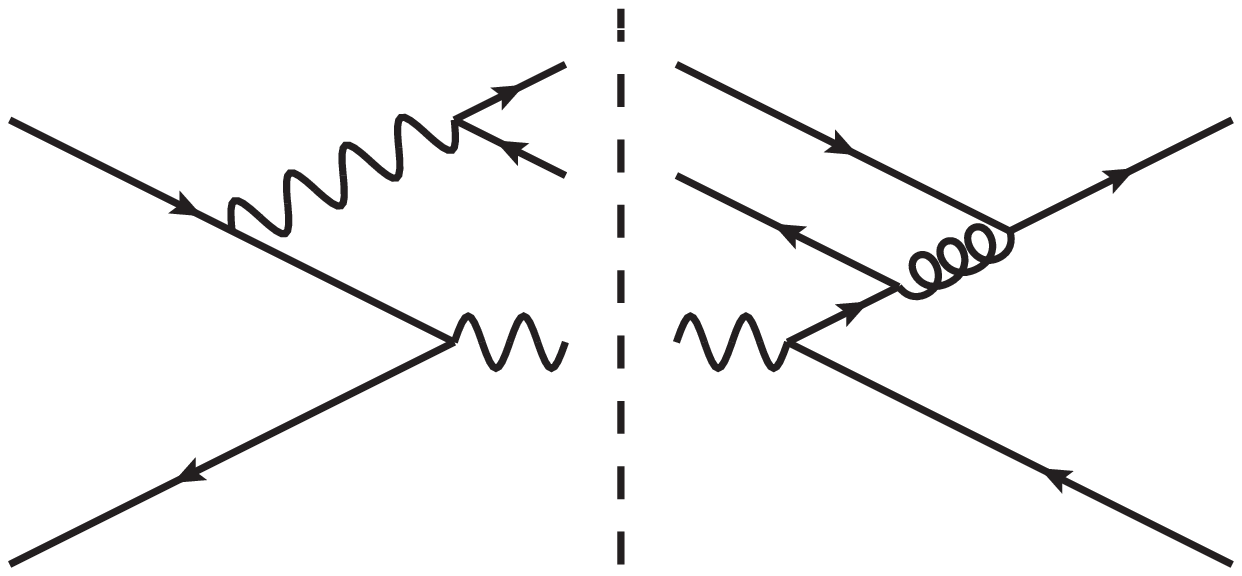}};
  \node at (3.2,-2.6) {,};
  \draw[->,black,thick] (-2.5,-2.4) -- (-1.5,-2.4);
 \end{tikzpicture}
 \caption{Contributions to the colour factor $C_F^2$ and their abelianised counterparts.}\label{vvrv}
\end{figure}

Consider the double-virtual corrections, shown on the top line of Fig.~\ref{vvrv}.
Upon setting $C_A$ to zero, colour traces that appear in both planar and non-planar diagrams
provide a colour factor $C_F^2$. Since any of the two gluon lines can be replaced by a photon line in any of these  diagrams, the
required modification of the colour factor is $C_F^2 \to 2 C_F e_q^2$.  It is easy to see that the same holds for
real-virtual contributions, shown on the second line of Fig.~\ref{vvrv}, 
and for interferences that arise between  double-real contributions (see the final line Fig.~\ref{vvrv}).

\begin{figure}[h]
\centering
 \begin{equation}
  \frac{1}{2}\left(\left|\adjustbox{valign=c}{\includegraphics[width=0.2\textwidth]{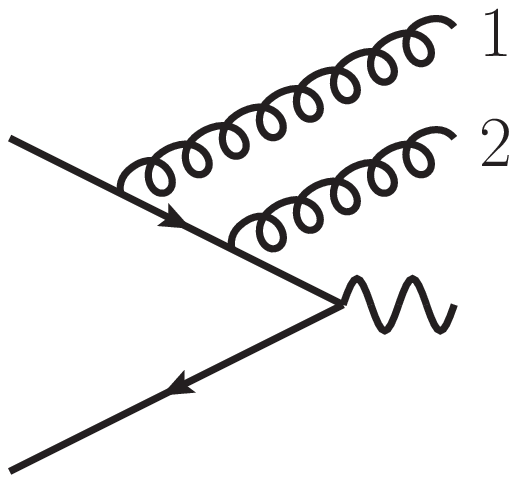}}\hspace{-12pt}+\adjustbox{valign=c}{\includegraphics[width=0.2\textwidth]{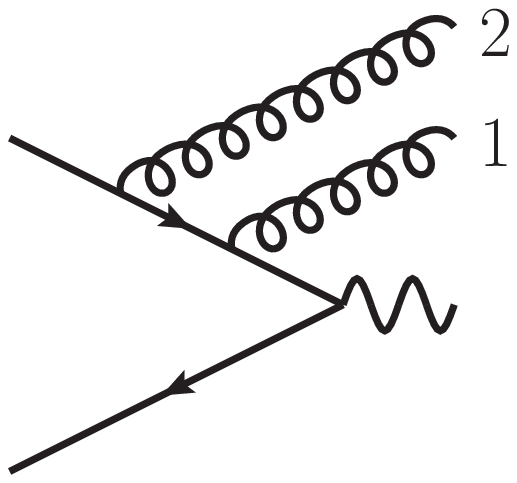}}\hspace{-12pt}\right|^2\right)\longrightarrow\left|\adjustbox{valign=c}{\includegraphics[width=0.15\textwidth]{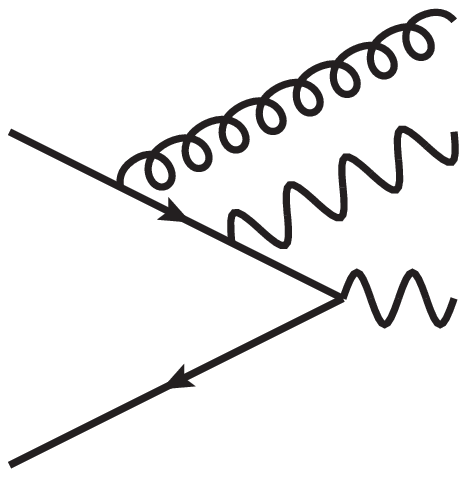}}\right|^2+\left|\adjustbox{valign=c}{\includegraphics[width=0.15\textwidth]{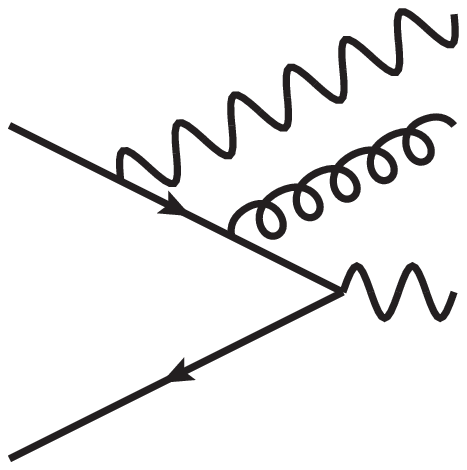}}\right|^2\nonumber
 \end{equation}
 \caption{Contributions to the colour factor $C_F^2$ and their abelianised counterparts.}\label{qqbgg}
\end{figure}

A distinct situation arises in cases when two gluons appear in the final state, shown in Fig.~\ref{qqbgg}.
In this case, two diagrams in the QCD
case are mapped onto two diagrams in the QCD$\otimes$QED case; hence, it appears at first sight that for these diagrams
the correct replacement rule is $C_F^2 \to C_F e_q^2$, so that the factor of two is missing.
However, this is not the case because
contributions of diagrams with two gluons to the cross section are multiplied by  a factor
$1/2!$ to account for the symmetric final state.
Clearly, there is no such factor in case of the $g + \gamma$ final state.
This mismatch is accounted for if the colour factor $C_F^2$ in the $q \bar q \to Z + g+g $ contribution 
is again replaced by $2 C_F e_q^2$, in accord  with what is needed for double-virtual and real-virtual contributions. 

Hence, after an examination of all the cases, we conclude
that  for processes with an incoming quark-antiquark pair or interference-like contributions with  two
identical quarks, we replace 
\begin{equation}
 C_F^2 \rightarrow2C_F e_q^2\ , \;\;\; T_R \to 0\ ,\;\;\;\; C_A \to 0\ ,
\label{eq:rep}
\end{equation}
in the formulas that describe NNLO QCD  corrections to $Z$ boson production and obtain results for mixed QCD$\otimes$QED corrections. 
In Eq.~(\ref{eq:rep}), $e_q$ is the electric charge of the incoming quark. 

For processes with an incoming (anti)quark and a gluon, there is no symmetry factor, and the two possible ways to replace a gluon by a photon amount to the two distinct processes $qg\rightarrow Z+q+\gamma$ and $q\gamma\rightarrow Z+q+g$.
The replacement rules become 
\begin{equation}
C_F^2 \rightarrow C_F e_q^2\ , \;\;\;  C_A \to 0\ .
\end{equation}
Similarly, for processes induced by two gluons, replacing a gluon by a photon leads to the processes $g\gamma\rightarrow Z+q\bar{q}$ and $\gamma g\rightarrow Z+q\bar{q}$. We then have the replacement rule
\begin{equation}
 C_F^2\rightarrow C_F e_q^2\ ,\;\;\; C_A \to 0\ .
\end{equation}
We also note that, if  photon-induced contributions are obtained from gluon-induced processes, the 
averaging over colour charges of the incoming partons  has to be changed as well.

Making use of the procedure described above, we abelianised the fully-differential description of the on-shell $Z$ boson production given in Ref.~\cite{Caola:2019nzf}, including regulated double-real and real-virtual contributions, integrated subtraction terms
and the  virtual contributions.
This gives us an opportunity to compute mixed QCD$\otimes$QED corrections to any infrared-safe observable in the production of an on-shell $Z$ boson.

\subsection{ Implementation of $Z$ boson decay}\label{xsdef}
We now discuss the treatment of the decay of the $Z$ boson into massless leptons, $Z \to \ell^+ \ell^-$,
in QCD and QED perturbative expansions. 
At leading order,
the cross section for the production of an on-shell $Z$ boson is computed from the tree-level process $q\bar{q}\rightarrow Z\rightarrow \ell^+\ell^-$, where the square
of the propagator of an intermediate $Z$ boson  is replaced by its narrow width limit
\begin{equation}
\label{eq:NWA}
 \frac{1}{(Q^2-M_Z^2)^2+M_Z^2\Gamma_Z^2}\rightarrow\frac{\pi}{M_Z\Gamma_Z}\delta(Q^2-M_Z^2)\ .
\end{equation}
In Eq.~(\ref{eq:NWA}) $Q$ is the four-momentum of the $Z$ boson and $\Gamma_Z$ is its width.
In principle, the width of the $Z$ boson in Eq.~(\ref{eq:NWA})
receives perturbative corrections.
These corrections should, partially, cancel QED corrections to $Z\rightarrow \ell^+ \ell^-$ that are included in
our  calculation. In order to account for that, we rewrite the cross section for $pp\rightarrow Z\rightarrow \ell^+\ell^-$
as follows
\begin{align}
\label{eq:def_xs_full}
\DIFFL \sigma_{pp\rightarrow Z\rightarrow \ell^+\ell^-} = \frac{\DIFFL \sigma_{pp\rightarrow Z} \DIFFL \Gamma_{Z\rightarrow \ell^+\ell^-} }{\Gamma_Z} = \text{Br}_{Z\rightarrow \ell^+\ell^-} \times \DIFFL \sigma_{pp\rightarrow Z}  \times \frac{ \DIFFL \Gamma_{Z\rightarrow \ell^+\ell^-} }{\Gamma_{Z\rightarrow \ell^+\ell^-}} \ .
\end{align}
In Eq.~(\ref{eq:def_xs_full}) we introduced the branching ratio of the $Z$ boson decay to a massless $\ell^+\ell^-$ pair
$\text{Br}_{Z\rightarrow \ell^+\ell^-}$; we will  treat it as an experimental input
parameter and will not expand it in $\alpha_s$ and $\alpha$. 
However, all other terms in Eq.~(\ref{eq:def_xs_full}) will
be treated within QCD/QED perturbation theory. In particular, the ratio
$\DIFFL \Gamma_{Z\rightarrow \ell^+\ell^-} / \Gamma_{Z\rightarrow \ell^+\ell^-}$
must be expanded to first order in $\alpha$. We write
\be
\begin{split}
 \Gamma_{Z\rightarrow \ell^+\ell^-}
=&  \Gamma_{Z\rightarrow \ell^+\ell^-}^{(0)} +  \Gamma_{Z\rightarrow \ell^+\ell^-}^{(1)} + {\cal O}(\alpha^2)\ , \\
\DIFFL \Gamma_{Z\rightarrow \ell^+\ell^-}
=& \DIFFL \Gamma_{Z\rightarrow \ell^+\ell^-}^{(0)} + \DIFFL \Gamma_{Z\rightarrow \ell^+\ell^-}^{(1)} + {\cal O}(\alpha^2)\ ,
\end{split}
\ee
and expand the ratio
\begin{equation}
\frac{ \DIFFL \Gamma_{Z\rightarrow \ell^+\ell^-} }{\Gamma_{Z\rightarrow \ell^+\ell^-}} =\frac{\DIFFL \Gamma_{Z\rightarrow \ell^+\ell^-}^{(0)}}{ \Gamma_{Z\rightarrow \ell^+\ell^-}^{(0)}}+\left[-\frac{\DIFFL \Gamma_{Z\rightarrow \ell^+\ell^-}^{(0)} \Gamma_{Z\rightarrow \ell^+\ell^-}^{(1)}}{\left(\Gamma_{Z\rightarrow \ell^+\ell^-}^{(0)}\right)^2}  + \frac{\DIFFL \Gamma_{Z\rightarrow \ell^+\ell^-}^{(1)}}{ \Gamma_{Z\rightarrow \ell^+\ell^-}^{(0)}}\right]+\mathcal{O}(\alpha^2)\ .
\end{equation}
By construction, the above expression integrates to one
over the unrestricted decay phase-space, so that terms in the square brackets integrate to zero.
The expansion coefficients of the leptonic $Z$ width read 
\begin{align}
 \Gamma_{Z\rightarrow \ell^+\ell^-}^{(0)}&=\frac{G_F M_Z^3}{6\sqrt{2}\pi}\left(\frac{1}{4}+\left(\frac{1}{2}-2\sin^2\theta_W\right)^2\right)\ ,\\
 \Gamma_{Z\rightarrow \ell^+\ell^-}^{(1)}&=\Gamma_{Z\rightarrow e^+e^-}^{(0)} \times \frac{3\alpha}{4\pi}\ . \
\end{align}

We now go back to Eq.~(\ref{eq:def_xs_full}) and expand all relevant ingredients in series in the strong and electromagnetic
coupling constants.  To present the
results of such an expansion, we
denote  the $\mathcal{O}(\alpha_s^n\; \alpha^m)$ contribution to
the $Z$ production cross section as $\sigma_{pp\rightarrow Z}^{(n,m)}$, and find the
following results  for  $pp \to Z \to \ell^+\ell^-$
cross sections\footnote{From now on we drop the subscripts in $\Gamma_{Z\rightarrow \ell^+\ell^-}$ and $\DIFFL\Gamma_{Z\rightarrow \ell^+\ell^-}$.}
\begin{align}
\label{eq:xs_LO} \DIFFL \sigma_{\mathrm{LO}}^{\phantom{()}} & = \text{Br}_{Z\rightarrow \ell^+\ell^-} \times \DIFFL \sigma^{(0,0)}_{pp\rightarrow Z}  \times  \frac{\DIFFL \Gamma^{(0)}}{ \Gamma^{(0)}}\ , \\
\label{eq:xs_NLO10} \DIFFL \sigma^{(\alpha_s)}_{\mathrm{NLO}} &  = \text{Br}_{Z\rightarrow \ell^+\ell^-} \times \DIFFL \sigma^{(1,0)}_{pp\rightarrow Z}  \times  \frac{\DIFFL \Gamma^{(0)}}{ \Gamma^{(0)}}\ , \\
\label{eq:xs_NLO01} \DIFFL \sigma^{(\alpha)}_{\mathrm{NLO}} &  = \text{Br}_{Z\rightarrow \ell^+\ell^-} \times \left [  \DIFFL \sigma^{(0,1)}_{pp\rightarrow Z}  \times  \frac{\DIFFL \Gamma^{(0)}}{ \Gamma^{(0)}}  + \DIFFL \sigma^{(0,0)}_{pp\rightarrow Z}  \times \left( \frac{\DIFFL \Gamma^{(1)}}{ \Gamma^{(0)}} - \frac{\DIFFL \Gamma^{(0)}}{ \Gamma^{(0)}} \times \frac{3\alpha}{4\pi}  \right) \right ] \ , \\
\label{eq:xs_NNLO20}\DIFFL \sigma^{(\alpha_s^2)}_{\mathrm{NNLO}} &  = \text{Br}_{Z\rightarrow \ell^+\ell^-} \times \DIFFL \sigma^{(2,0)}_{pp\rightarrow Z}  \times  \frac{\DIFFL \Gamma^{(0)}}{ \Gamma^{(0)}} \ , \\
\label{eq:xs_NNLO11}\DIFFL \sigma^{(\alpha_s\alpha)}_{\mathrm{NNLO}} &  = \text{Br}_{Z\rightarrow \ell^+\ell^-} \times \left [ \DIFFL \sigma^{(1,1)}_{pp\rightarrow Z}  \times \frac{\DIFFL \Gamma^{(0)}}{ \Gamma^{(0)}}  + \DIFFL \sigma^{(1,0)}_{pp\rightarrow Z}  \times \left( \frac{\DIFFL \Gamma^{(1)}}{ \Gamma^{(0)}} - \frac{\DIFFL \Gamma^{(0)}}{ \Gamma^{(0)}} \times \frac{3\alpha}{4\pi}  \right)  \right ]\ .
\end{align}
Note that terms in round brackets in Eqs.~(\ref{eq:xs_NLO01}) and (\ref{eq:xs_NNLO11})  integrate
to zero over an unrestricted phase-space, such that the inclusive cross section is given by a product of the branching fraction and the production cross section, 
as expected from Eq.~(\ref{eq:def_xs_full}).
We emphasise again that we consider massless leptons throughout this paper.

To present our results for QCD$\otimes$QED corrections, we define ratios of contributions to 
the cross section at different perturbative orders.
We write 
\begin{align}
\label{eq:kfactorsInc}
\Delta_{\alpha} = \frac{\sigma^{(\alpha)}_{\mathrm{NLO}}}{\sigma_{\mathrm{LO}}^{\phantom{()}}+\sigma^{(\alpha_s)}_{\mathrm{NLO}}}\ ,\;\;\;\;\;\;
\Delta_{\alpha_s^2 }
= \frac{\sigma^{(\alpha_s^2)}_{\mathrm{NNLO}}}{\sigma_{\mathrm{LO}}^{\phantom{()}}+\sigma^{(\alpha_s)}_{\mathrm{NLO}}}\ ,\;\;\;\;\;\;
\Delta_{\alpha_s \alpha }
= \frac{\sigma^{(\alpha_s \alpha)}_{\mathrm{NNLO}}}{\sigma_{\mathrm{LO}}^{\phantom{()}}+\sigma^{(\alpha_s)}_{\mathrm{NLO}}}\ .
\end{align}
To discuss kinematic distributions, we define differential bin-by-bin corrections in a similar fashion 
\be
  \DIFFL \Delta_{\alpha} =  \frac{{\rm d} \sigma^{(\alpha)}_{\mathrm{NLO}}}{{\rm d}
    \sigma_{\mathrm{LO}}^{\phantom{()}}+{\rm d} \sigma^{(\alpha_s)}_{\mathrm{NLO}}}\ , \;\;
\DIFFL \Delta_{\alpha_s^2}
=  \frac{{\rm d} \sigma^{(\alpha_s^2)}_{\mathrm{NNLO}}}{{\rm d} \sigma_{\mathrm{LO}}^{\phantom{()}}+{\rm d} \sigma^{(\alpha_s)}_{\mathrm{NLO}}}\ ,\;\;
\DIFFL \Delta_{\alpha_s \alpha }
=  \frac{{\rm d} \sigma^{(\alpha_s \alpha)}_{\mathrm{NNLO}}}{{\rm d} \sigma_{\mathrm{LO}}^{\phantom{()}}+{\rm d} \sigma^{(\alpha_s)}_{\mathrm{NLO}}}\ .
\ee

\subsection{Checks of the computation}\label{checks}

Although the abelianisation procedure is, in principle, straightforward, its implementation in a fully-differential NNLO QCD
computation is tedious. For this reason, it is important to check the implementation. We did this in
the following way. In addition to abelianising the fully-differential NNLO QCD computation in Ref.~\cite{Caola:2019nzf},
we also abelianised the analytic NNLO QCD coefficients for inclusive Z boson production
given in Ref.~\cite{Hamberg:1990np} and compared these 
  to  analytic expressions for mixed QCD$\otimes$QED corrections
  given  in appendix B  of Ref.~\cite{deFlorian:2018wcj}. We then used
  our abelianised analytic expressions to compute the inclusive on-shell production cross section of $pp \to Z$
  and check that it agrees with the  cross section  obtained using the fully-differential implementation of mixed
  QCD$\otimes$QED corrections.

%% file: sections/03_results.tex
\section{Results}\label{results}

In this section we  discuss mixed QCD$\otimes$QED corrections to various observables in $pp \to Z \to \ell^+\ell^-$
 and compare them to other corrections.
We consider the LHC with 13 TeV center-of-mass collision energy. We use  $M_Z=91.1876~\gev$ for the mass of the $Z$ boson,
and consider its decay to a single flavour of massless leptons with a branching ratio $\text{Br}_{Z\rightarrow \ell^+\ell^-}=0.033632$.
We compute the couplings of the $Z$ boson to leptons and quarks using $G_F=1.16639\times10^{-5}~\gev^{-2}$,
$M_W=80.398~\gev$ and  $\sin^2\theta_W=0.2226459$ as the  input parameters. 
We use the NNPDF3.1luxQED set with five active flavours \cite{Bertone:2017bme} as provided by the LHAPDF interface \cite{Buckley:2014ana},
and take all quark to be massless.
To compute the  $\Delta$-corrections defined in the previous Section, we always use parton distributions
at  NNLO accuracy to calculate  all the relevant contributions.
We set the renormalisation and factorisation scales to $\mu_R=\mu_F=M_Z$.
The strong coupling constant is taken to be $\as(M_Z)=0.118$ which is compatible with values provided by the NNPDF set.
We describe  photon interactions with 
leptons  and quarks by the fine structure constant evaluated at the renormalisation
scale $\mu = M_Z$; numerically, it is equal to    $\alpha = 1/128$. 

\subsection{Inclusive cross sections}

We use the above setup to compute the inclusive cross section of $Z$ boson production at various (NLO QED,
NNLO QCD and mixed QCD$\otimes$QED)  approximations. Using notations introduced in the previous section, we  find\footnote{
  We neglect contributions of top quarks to $Z$-boson  production cross section including  the
  top-bottom triangle correction to the axial current. Such contributions were  shown to be small
  in Ref.~\cite{Dicus:1985wx}.}
\begin{equation}
  \Delta_{\alpha}=3.2\cdot10^{-3} \ ,\qquad \Delta_{\alpha_s^2}=-6.4\cdot10^{-3}\ ,\qquad \Delta_{\alpha_s\alpha}= 2.9\cdot10^{-4} \ .
  \label{eq3.3}
\end{equation}
We note that the $\Delta$ ratios in Eq.~(\ref{eq3.3}) receive contributions  from corrections to the production only, 
since corrections to the decay of the $Z$ boson $Z \to \ell^+\ell^-$
cancel with  corrections to the partial decay  width $\Gamma_{Z \to \ell^+\ell^-}$, as explained in the previous Section.

It follows from Eq.~(\ref{eq3.3})  that  the magnitude of mixed QCD$\otimes$QED corrections to the
inclusive cross section $pp \to Z$ is consistent with  
expectations. Indeed,  they are smaller than NLO QED (NNLO QCD) corrections by a factor ten (twenty), respectively.
These suppression factors are in accord  with our expectations,  based on the
relative magnitude of strong and electromagnetic coupling constants.   

We note that our results for the corrections to the inclusive cross section
are different from the results of Ref.~\cite{deFlorian:2018wcj}.
In particular, this reference reported a smaller suppression of the mixed QCD$\otimes$QED corrections relative
to the NNLO QCD one.
The reason for this is that Ref.~\cite{deFlorian:2018wcj} employs a four-flavour scheme to compute both NNLO QCD and
mixed QCD$\otimes$QED corrections. 
Since both 
NNLO QCD and mixed QCD$\otimes$QED corrections exhibit a strong sensitivity to input parameters, 
thanks to a very strong cancellation between
(large) corrections to $q \bar q$ and $q g$ partonic channels at the LHC energies~\cite{deFlorian:2018wcj},
even small changes in the input can lead to significant changes in final results for corrections. 
We have confirmed that if we use the same input,
we agree with the results of Ref.~\cite{deFlorian:2018wcj}.\footnote{We thank D. de Florian for clarifications and  help with this comparison.}

Finally, we comment on the
scale dependence of the total NNLO cross section that  includes  the $\Delta_{\alpha_s^2}$ and $\Delta_{\alpha_s \alpha}$ corrections.
Although this dependence is small, ${\cal O}(0.5-0.7 \%)$,  it  is entirely dominated by pure QCD effects and it is not possible
to unambigously identify the impact of QCD$\otimes$QED corrections on it.
For this reason, we decided to avoid presenting results for the scale dependence of the NNLO cross section.
Instead, to understand the dependence of the QCD$\otimes$QED corrections on  the scale choice,
we  consider the $\Delta_{\alpha_s \alpha}$ correction where the cancellation  between the $\mu$-dependence of parton
distribution functions and the explicit  scale dependence of the NNLO contribution cannot be expected.  Because of that,
it is not
surprising  that these corrections appear to be
rather sensitive to the choice of the factorisation and the renormalisation scales $\mu$. Indeed, by choosing the scale
$\mu = \left [ M_Z/2,M_Z,2 M_Z \right ]$,
we obtain the corrections $\Delta_{\alpha_s \alpha}  = \left [ 5.6 , 2.9 , -0.28  \right ] \cdot 10^{-4}$.
Although a stronger sensitivity of the {\it correction} $\Delta_{\alpha_s \alpha}$
to the choice of the scale is expected, there is yet another reason for large variations in $\Delta_{\alpha_s \alpha}$.
In fact, the enhanced dependence on $\mu$  can be also traced back to a strong cancellation between
quark- and gluon-initiated contributions to $\Delta_{\alpha_s \alpha}$ which is affected by the change of the scale $\mu$ in a significant way.

As we discuss in the next Section, the cancellation between $q \bar q$ and $q g$ channels is also
an important feature of mixed corrections to fiducial cross sections.
We therefore expect that for the  fiducial cases the scale variation of mixed corrections will exhibit  similar behaviour.
For this reason, we will not discuss the scale dependences of those corrections  any further in the next section. 

\subsection{Fiducial cross sections}

Fiducial cross sections are defined through kinematic selection criteria applied to  physical objects in final states.
We define the selection criteria for $pp \to Z \to \ell^+ \ell^-$ as
\begin{equation}
  p_{\perp,\ell_1}>24~\mathrm{GeV}\ ,\quad p_{\perp,\ell_2}>16~\mathrm{GeV}\ ,\quad|y_{\ell_i}|<2.4\ ,\;\;\;\;
   50~\mathrm{GeV} < m_{\ell\bar{\ell}} < 120~{\rm GeV}\ ,
  \label{eq:sc}
\end{equation}
where $\ell_{1,2}$ denote leptons with leading and subleading transverse momenta, respectively.

Since we work with massless leptons, their  transverse momenta are  not collinear-safe observables; for this reason,
we need to introduce an analog of QCD jets for leptons by combining leptons with collinear photons. 
Such recombination procedures are also used in experimental measurements to define ``physical'' electrons
subject to  selection cuts.

For the purposes of the computation of mixed QCD$\otimes$QED corrections,
we choose a simplified version of the standard recipe \cite{Alioli:2016fum}. To this end, we begin by 
computing two quantities  $R_{\ell^{\pm}\gamma}=\sqrt{(y_{\ell^\pm}-y_{\gamma})^2+(\varphi_{\ell^{\pm}}-\varphi_{\gamma})^2}$ 
where $y_{\ell^\pm,\gamma}$ are  the rapidities  and $\varphi_{\ell^\pm,\gamma}$ the azimuthal angles of
the lepton or antilepton,  and the photon, respectively.
If  $R_{\ell^{\pm}\gamma}$ for the photon and one of the leptons is smaller than some $R_{\mathrm{min}}$, 
the photon is recombined with the lepton by adding their momenta; the new object is treated as a lepton inasmuch
as  the selection cuts Eq.~(\ref{eq:sc}) are concerned.  In this paper, we use the standard  value
 \cite{Alioli:2016fum}
$R_{\mathrm{min}}=0.1$.

As we already mentioned, there are three distinct sources of QED corrections. To show them separately, 
we  decompose the NLO QED and mixed QCD$\otimes$QED results according to whether the QED correction
is associated with the production of the $Z$ boson ($P$), its decay ($D$), or its decay width $\Gamma_{Z \to e^+e^-}$ ($W$). Using the definitions for
$\Delta$'s in the previous Section,  we find
\begin{equation}
\begin{split}
& \Delta_{\alpha}=(3.0\cdot10^{-3})_{P}-(7.2\cdot10^{-3})_{D}-(1.6\cdot10^{-3})_{W}\ , \\
& \Delta_{\alpha_s^2}=-(1.2\cdot10^{-2})\ , \\
& \Delta_{\alpha_s\alpha}=-(1.5\cdot10^{-4})_{P\otimes P}-(4.9\cdot10^{-3})_{P\otimes D}-(0.3\cdot10^{-3})_{P \otimes W}\ .
  \label{eq101}
  \end{split} 
\end{equation}

The many
results in Eq.~(\ref{eq101}) can be compared in different ways. First, we note that the QCD corrections are larger than in the
inclusive case by almost a factor of two whereas the NLO QED
corrections (initial)  do not change significantly.
The mixed QCD$\otimes$QED correction to the
production is, on the other hand, smaller by a factor of two than in the inclusive case.
Similar to the inclusive case, the relative
magnitude of NLO QED, NNLO QCD and mixed QCD$\otimes$QED  corrections to the  fiducial $Z$ boson production cross section is consistent
with expectations based on the relative sizes of QCD and QED couplings,
despite the somewhat larger relative magnitude of the NNLO QCD correction.

It is interesting to understand how the final result for QCD$\otimes$QED  corrections to the  production  comes about. To this end, 
it is instructive to  decompose $[\Delta_{\alpha_s\alpha}]_{P\otimes P}$ into contributions of particular partonic channels, see
Table~\ref{table:channels}.
We observe a sizeable, almost an order-of-magnitude   cancellation between $q \bar q$ and $qg $ channels.
In fact a similar cancellation reduces the magnitude of NNLO QCD corrections which could have been
quite a bit  bigger than what they are if this cancellation was not  present. 

It is seen from Table~\ref{table:channels} that contributions of photon-induced channels  are
very small, as expected. However, due to the aforementioned cancellation, they still
contribute roughly twenty percent to the total  result for mixed initial-initial corrections. It is also interesting that 
the photon-induced contributions are  {\it larger} than
those of $qq$ channels.  This implies that neglecting  contributions with photons in the initial state is not a
good approximation if QCD$\otimes$QED precision is desired.

\begin{table}[t]
 \centering
 \begin{tabular}{|r|c|}
   \hline
  Partonic Channel &  $[\Delta_{\alpha_s\alpha}]_{P \otimes P} \cdot 10^{4}$ \\
  \hline $q\bar{q}$ & 5.60 \\
  $qq$ & 0.13 \\
  $qg + gq$ & -7.01 \\
  $q\gamma + \gamma q$ & -0.32 \\
  $\gamma g$ & 0.06 \\
  ${\rm Total}$ & -1.54 \\
  \hline
 \end{tabular}
 \caption{Contributions of the different partonic  channels to $[\Delta_{\alpha_s\alpha}]_{P \otimes P}$.}
  \label{table:channels}
\end{table}

A rather different situation arises if we look at contributions to Eq.~(\ref{eq101})  that involve 
corrections to $Z$ boson decays. They are described by $[\Delta_{\alpha}]_D$  and by $[\Delta_{\alpha_s\alpha}]_{P \otimes D}$ for the
QED and QCD$\otimes$QED corrections, respectively.  Inspecting  Eq.~(\ref{eq101}), we observe that both of these contributions
are large and that $[\Delta_{\alpha_s \alpha}]_{P \otimes D}$  is smaller than   $[\Delta_{\alpha}]_D$ by only thirty percent
in spite of being  suppressed by one power of $\alpha_s$.  This implies that for
fiducial cross sections   QED radiation in the decay is very strongly affected by QCD radiation in the production. 

This result illustrates that the selection criteria shown
in Eq.~(\ref{eq:sc}) are strongly impacted by the non-vanishing transverse 
momentum of the $Z$ boson that in the case of a fixed-order computation  is provided by the
initial state QCD radiation. In addition, $ R_{\min}=0.1$
is probably too small an isolation cone to allow
a stable perturbative description of QED radiation off  the outgoing leptons. To illustrate this remark, we 
point out that,
with $R_{\rm min} =0.1$ and an additional selection cut $p_{\perp, \gamma} > 5~{\rm GeV}$, the rate of the
$Z$ boson decay to
three QED jets becomes close to
four percent  and thus much larger than a naive expectation based on $\alpha/\pi \sim 2 \times 10^{-3}$ suppression
of events with additional radiation. It is clear that a quasi-collinear fragmentation of a lepton to a photon,
allowed by the selection cuts, may strongly change the observable cross section by reducing the transverse momentum of the lepton. 
Furthermore, we also observe that by rejecting events with two leptons and a
photon and keeping events with only
Born-like kinematics,  the size of  $[\Delta_{\alpha_s\alpha}]_{P \otimes D}$ gets reduced relative to $[\Delta_{\alpha}]_D$.\footnote{
  In the future, it may  be interesting 
  to combine  $\mathcal{O}(\alpha_s^2)$ corrections to the production of the $Z$ boson and $\mathcal{O}(\alpha)$ to its
decay in order to investigate the size of the correction due to a second emission from the initial state; since the first QCD emission
already provides some boost to the $Z$ boson, it is conceivable  that the corrections due to second gluon emission will be much more moderate.} 

\subsection{Kinematic distributions }

We continue with the discussion of the impact of QCD$\otimes$QED corrections
on kinematic distributions for the production of two leptons via an on-shell $Z$ boson at the LHC.
Below we show the respective distribution at NLO QCD accuracy in upper panels and 
the relative corrections $\Delta_{\alpha_s^2}$, $\left[ \Delta_{\alpha_s\alpha} \right ]_{P \otimes P}$
and $\left [ \Delta_{\alpha_s\alpha} \right ]_{P \otimes D}$ in lower panels. We use the  selection
criteria  shown in Eq.~(\ref{eq:sc}) and $R_{\rm min} = 0.1$  throughout this section. 

\begin{figure}[h]
\centering
\resizebox{0.48\textwidth}{!}{\input{figures/plots/DY_Histos_2108X_ptll}}
\resizebox{0.48\textwidth}{!}{\input{figures/plots/DY_Histos_2108X_yll}}
\caption{Relative differential corrections $\Delta$ for the transverse momentum (left) and rapidity (right) of the dilepton system, $p_{\perp,\ell \ell}$ and $y_{\ell \ell}$. See text for further details.}
\label{figure:llplots}
\end{figure}
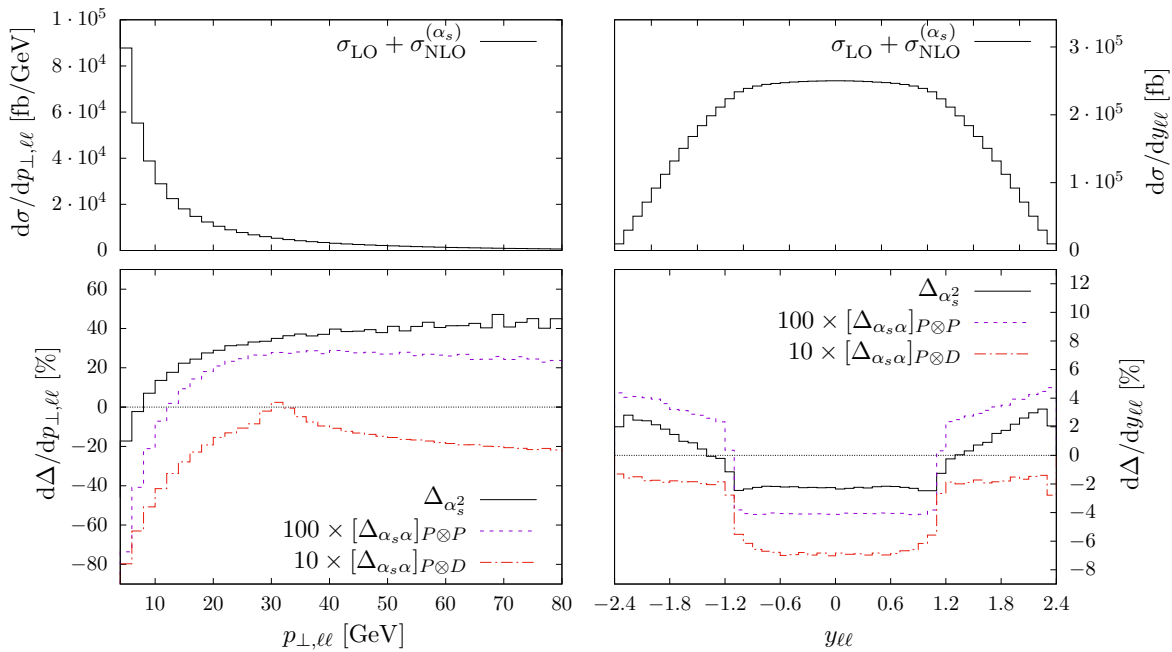

We begin with the discussion of the transverse momentum distribution of the two leptons,
$p_{\perp,\ell\ell}$, shown in the left panels of Fig.~\ref{figure:llplots}.
It is seen from the plot  that both
$\Delta_{\alpha_s^2}$ and $[\Delta_{\alpha_s\alpha}]_{P \otimes P}$ corrections
become flat and positive above $p_{\perp,\ell\ell} \sim 20~{\rm GeV}$.
In that  region, the NNLO QCD corrections amount to ${\cal O}(40)$ percent. Although this
is quite a large correction, we note 
that in this kinematic
region they can be considered as NLO QCD corrections to $Z+{\rm jet}$  production.
The mixed $[\Delta_{\alpha_s\alpha}]_{P \otimes P}$
corrections  can be thought of as  QED corrections to the NLO QCD distribution; this would suggest a percent-level correction.
In fact, initial-initial 
corrections in this case 
turn out to be even smaller, of the order of two permille for $p_{\mathrm{T,\ell\ell}}>20~$GeV. This is partially
due to the  cancellation between $q\bar{q}$- and $qg$-induced contributions that we already discussed.
For smaller values of the $Z$ transverse momentum, the corrections become large and negative;
however,  resummation may be needed to  obtain a reliable prediction in this region.  

The initial-final correction $[\Delta_{\alpha_s\alpha}]_{P \otimes D}$ to the $p_{\perp,\ell\ell}$ distribution of a lepton pair 
shows quite  a different behaviour, with a maximum at $p_{\perp,\ell\ell}\simeq 30~$GeV. This feature appears
because of an interplay of a few  contributions with different kinematics features.
On the one hand, processes without initial-state QCD radiation
but with 
final-state photon emission yield a pair of leptons with a  total transverse momentum  smaller than $M_Z/2$.
Moreover, the selection cuts we use further restrict it to $p_{\perp,\ell\ell}\lesssim~31~$GeV. On the other hand, processes with initial state radiation boost the $Z$ boson, leading to a tail which extends beyond this kinematic limit. This behaviour can already be observed in the initial and final contributions at NLO QED accuracy.

It is also interesting to point out that, although initial-final $[\Delta_{\alpha_s\alpha}]_{P \otimes D}$
corrections are indeed larger
than initial-initial corrections $[\Delta_{\alpha_s\alpha}]_{P \otimes P}$ for
almost all values of the transverse momentum $p_{\perp,\ell\ell}$, it is not the case
for $p_{\perp,\ell\ell} \sim ~30~{\rm GeV}$ where the initial-final correction passes
through zero(s). For those values of $p_{\perp,\ell \ell}$, the  initial-final and initial-initial
QCD$\otimes$QED contributions become comparable. 

The rapidity distribution of the dilepton system at NLO QCD  and the different corrections to it are 
displayed in the right panels of Fig.~\ref{figure:llplots}. 
The shape of the rapidity distribution is  determined by the selection cuts which   flatten the distribution for
values $|y_{\ell\ell}|<1$. Inside this region, which represents the bulk of the cross section, the NNLO QCD correction is
flat and negative and
amounts to a decrease of the NLO QCD result by  $\mathcal{O}(-2)$ percent. The NNLO QCD correction then
crosses zero at rapidities $|y_{\ell\ell}|\simeq 1$ and increase to $\mathcal{O}(3)$ percent at large rapidities.
The mixed $[\Delta_{\alpha_s\alpha}]_{P \otimes P}$ correction has a very similar shape; it  decreases the NLO QCD distribution by 
$\mathcal{O}(-0.4)$ permille in the central region and increases it  to $\mathcal{O}(+0.4)$ permille outside.
In contrast to the previous two corrections, the initial-final
$[\Delta_{\alpha_s\alpha}]_{P \otimes D}$ correction is negative for all rapidities and amounts to a decrease by about $\mathcal{O}(-7)$
permille in the central region. For rapidities $|y_{\ell\ell}|>1$, the initial-final correction becomes smaller but 
it remains 
 a factor 5 larger than the initial-initial one.
Hence, it follows that
for the rapidity distribution of a lepton pair,  the initial-final corrections always dominate over the initial-initial one. 

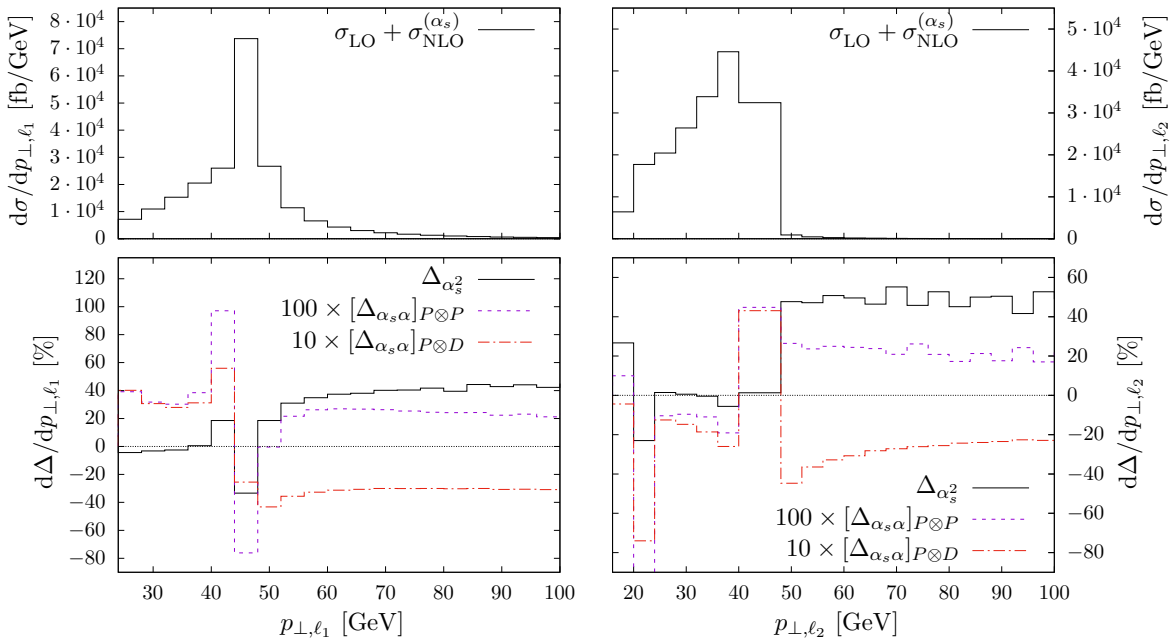
\begin{figure}[h]
\centering
\resizebox{0.48\textwidth}{!}{\input{figures/plots/DY_Histos_2108X_ptl1}}
\resizebox{0.48\textwidth}{!}{\input{figures/plots/DY_Histos_2108X_ptl2}}
\caption{Relative differential corrections $\Delta$ for the transverse momentum of the leading (left) and subleading (right) lepton,
  $p_{\perp,\ell_1}$ and $p_{\perp,\ell_2}$. See text for further details.}
\label{figure:ptlplots}
\end{figure}

We show the transverse momentum  distributions of the leading and the subleading leptons in Fig.~\ref{figure:ptlplots}. For the leading
lepton, the NNLO QCD corrections enhance the distribution at large $p_{\perp,\ell_1}$, which is consistent with the additional boost that
leptons get from the second initial-state emission. The feature at $p_{\perp,\ell_1}=M_Z/2$ is a Sudakov shoulder effect, which is known to appear close to kinematic boundaries. The NNLO QCD
correction factor stabilises at ${\cal O}(40)$ percent for large values of $p_{\perp,\ell1}$. The initial-initial
correction $[\Delta_{\alpha_s\alpha}]_{P \otimes P}$ shows a similar behaviour at high $p_{\perp,\ell_1}$ where it is about two permille. 
Interestingly, the initial-initial correction
also enhances the distribution at smaller values of the transverse momentum $p_{\perp,\ell_1}$. 
The initial-final correction $[\Delta_{\alpha_s\alpha}]_{P \otimes D}$ is
negative at high and positive at small $p_{\perp,\ell_1}$; this is consistent with the picture
of the final state leptons losing energy to  QED radiation and decreasing their transverse momenta. 
The correction $[\Delta_{\alpha_s\alpha}]_{P \otimes D}$ changes from ${\cal O}(-3)$ percent for  $p_{\perp,\ell_1}>M_Z/2$
to ${\cal O}(+3)$ percent for $p_{\perp,\ell_1}<M_Z/2$.

The transverse momentum distribution of the subleading lepton at NLO QCD
accuracy has some features which impact the respective corrections.
Indeed, in addition to the Sudakov shoulder at $p_{\perp,\ell_2}=M_Z/2$, the distribution also features a similar effect
at $p_{\perp,\ell_2} = 24~{\rm GeV}$, due to the fact that $24~{\rm GeV}$ is
a cut on the minimal transverse momentum of the leading lepton. 
Since at leading order the transverse momenta of the two leptons must be
equal to each other, the leading order $p_{\perp,\ell_2}$ distribution
is truncated at this value {\it also for the subleading lepton} and
the  lower bins are only populated through higher order  corrections.

In order to avoid displaying  large fluctuations of radiative effects around the Sudakov shoulder at $p_{\perp,\ell_2}=M_Z/2$, we
combined   two bins between $40~{\rm GeV}$ and $48~$GeV into a single bin to present  various corrections. 
Similar to the leading-lepton case, 
we observe large positive NNLO QCD and mixed $[\Delta_{\alpha_s\alpha}]_{P\otimes D}$ corrections for $p_{\perp,\ell_2} > M_Z/2$ that
can be as large as $\mathcal{O}(50)$ percent
for QCD and $\mathcal{O}(2)$ permille for mixed QCD$\otimes$QED. The initial-final correction $[\Delta_{\alpha_s\alpha}]_{P\otimes D}$
 is negative  and takes values between ${\cal O}(-2)$ and ${\cal O}(-4)$ percent.

 Finally,  we discuss distributions of  $\cos \theta^* $, where $\theta^*$ is the angle between the three-momentum
 of one of the leptons  and a unit vector constructed from the difference between three-momenta of the colliding
 protons 
 in the rest frame of the dilepton system. Its cosine is given by the following formula \cite{Collins:1977iv}
\begin{align}
\cos \theta^* = \frac{  P_{\ell^-}^+P_{\ell^+}^- - P_{\ell^-}^-P_{\ell^+}^+}{\sqrt{m^2_{\ell^+\ell^-}(m^2_{\ell^+\ell^-}+p_{\perp,\ell^+\ell^-})}}{} \frac{p_{z,\ell^+\ell^-}}{\left| p_{z,\ell^+\ell^-} \right|} \ ,
\end{align}
where $P_i^{\pm} = (E_i \pm p_{z,i})$. The  $\cos \theta^*$-distribution is one of the few observables with strong 
sensitivity  to the weak mixing angle $\sin^2\theta^{\ell}_{\rm eff}$; it is used in experimental analyses for this purpose
\cite{Sirunyan:2018swq}.   The relevant input for this measurement is provided by   $\cos \theta^*$ distributions
for restricted  $m_{\ell\ell}$ and $|y_{\ell\ell}|$ intervals.

\begin{figure}[htbp]
\centering
\resizebox{0.48\textwidth}{!}{\input{figures/plots/DY_Histos_2108X_costh2}}
\resizebox{0.48\textwidth}{!}{\input{figures/plots/DY_Histos_2108X_costh4}}
\caption{Relative differential corrections $\Delta$ for the $\cos \theta^*$ distribution for $0.6<|y_{\ell\ell}|<1.2$ (left)
  and $1.8<|y_{\ell\ell}|<2.4$ (right). See text for further details.}
\label{figure:costhplots}
\end{figure}
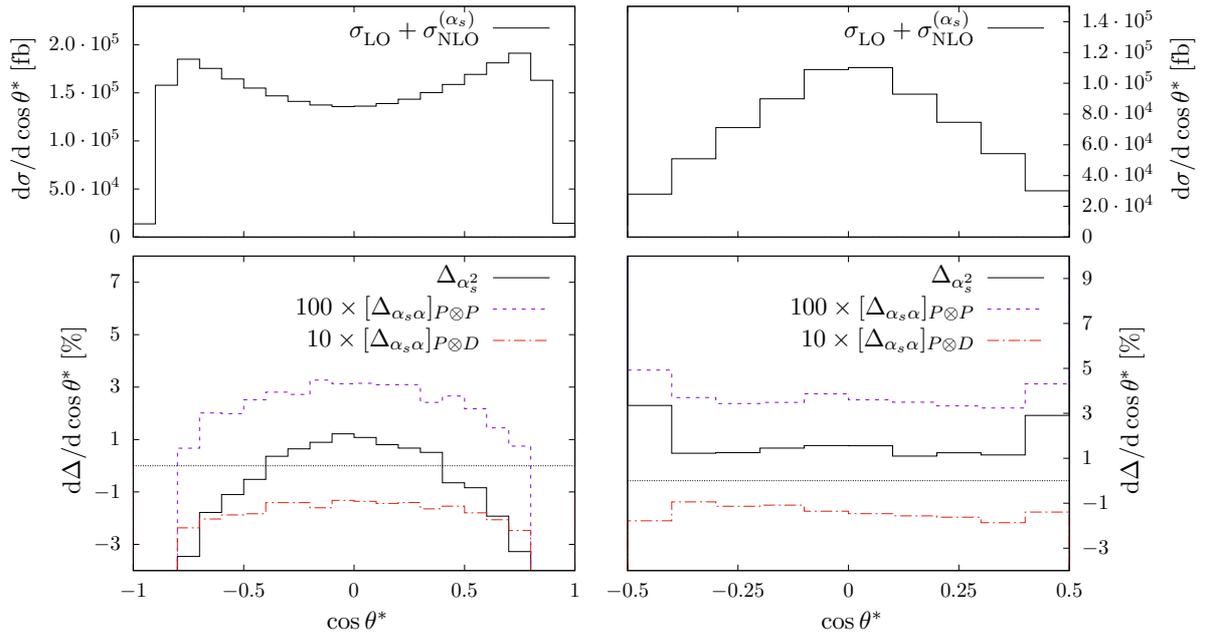

To illustrate how various effects modify the $\cos \theta^*$ distributions, 
in Fig.~\ref{figure:costhplots} we show them for  $0.6<|y_{\ell\ell}|<1.2$ and
$1.8<|y_{\ell\ell}|<2.4$ rapidity intervals. 
The panels on the left describe the rapidity interval $0.6<|y_{\ell\ell}|<1.2$ and feature 
 the distribution at NLO QCD accuracy   with
 two well-separated maxima in the upper panel and   NNLO QCD and mixed QCD$\otimes$QED
 corrections in the lower panel. 
 The NNLO QCD corrections are  below a  percent level at small values of $\cos\theta^*$, but become large
 and negative at the boundaries of the distribution.
 The mixed $[\Delta_{\alpha_s\alpha}]_{P \otimes P}$ corrections are relatively
 flat and amount to roughly $0.3$ permille in the bulk of the
 distribution. They   increase slightly at 
 small values of $\cos \theta^* $, but become large and negative in the outermost non-vanishing bins.
The initial-final  $[\Delta_{\alpha_s\alpha}]_{P \otimes D}$ contribution is negative for all values of $\cos \theta^* $
and is relatively flat, taking values between one and two permille.

The $\cos \theta^* $ distribution for rapidities $1.8<|y_{\ell\ell}|<2.4$ is quite different.
First, two maxima merge into 
one maximum located at small values  of $\cos \theta^* $. For this reason, we only display results
in the interval $-0.5 < \cos \theta^* < 0.5$.
Second, all  corrections in this case
are rather flat.
However, their magnitudes are comparable to those in the interval  $0.6<|y_{\ell\ell}|<1.2$,
with the NNLO QCD corrections being just few percent and the initial-initial
correction $[\Delta_{\alpha_s\alpha}]_{P \otimes P}$
below a permille level. The initial-final $[\Delta_{\alpha_s\alpha}]_{P \otimes D}$ corrections are negative and
are close to two permille.

%% file: figures/plots/DY_Histos_2108X_ptll.tex
\begingroup
  \makeatletter
  \providecommand\color[2][]{%
    \GenericError{(gnuplot) \space\space\space\@spaces}{%
      Package color not loaded in conjunction with
      terminal option `colourtext'%
    }{See the gnuplot documentation for explanation.%
    }{Either use 'blacktext' in gnuplot or load the package
      color.sty in LaTeX.}%
    \renewcommand\color[2][]{}%
  }%
  \providecommand\includegraphics[2][]{%
    \GenericError{(gnuplot) \space\space\space\@spaces}{%
      Package graphicx or graphics not loaded%
    }{See the gnuplot documentation for explanation.%
    }{The gnuplot epslatex terminal needs graphicx.sty or graphics.sty.}%
    \renewcommand\includegraphics[2][]{}%
  }%
  \providecommand\rotatebox[2]{#2}%
  \@ifundefined{ifGPcolor}{%
    \newif\ifGPcolor
    \GPcolortrue
  }{}%
  \@ifundefined{ifGPblacktext}{%
    \newif\ifGPblacktext
    \GPblacktexttrue
  }{}%
  \let\gplgaddtomacro\g@addto@macro
  \gdef\gplbacktext{}%
  \gdef\gplfronttext{}%
  \makeatother
  \ifGPblacktext
    \def\colorrgb#1{}%
    \def\colorgray#1{}%
  \else
    \ifGPcolor
      \def\colorrgb#1{\color[rgb]{#1}}%
      \def\colorgray#1{\color[gray]{#1}}%
      \expandafter\def\csname LTw\endcsname{\color{white}}%
      \expandafter\def\csname LTb\endcsname{\color{black}}%
      \expandafter\def\csname LTa\endcsname{\color{black}}%
      \expandafter\def\csname LT0\endcsname{\color[rgb]{1,0,0}}%
      \expandafter\def\csname LT1\endcsname{\color[rgb]{0,1,0}}%
      \expandafter\def\csname LT2\endcsname{\color[rgb]{0,0,1}}%
      \expandafter\def\csname LT3\endcsname{\color[rgb]{1,0,1}}%
      \expandafter\def\csname LT4\endcsname{\color[rgb]{0,1,1}}%
      \expandafter\def\csname LT5\endcsname{\color[rgb]{1,1,0}}%
      \expandafter\def\csname LT6\endcsname{\color[rgb]{0,0,0}}%
      \expandafter\def\csname LT7\endcsname{\color[rgb]{1,0.3,0}}%
      \expandafter\def\csname LT8\endcsname{\color[rgb]{0.5,0.5,0.5}}%
    \else
      \def\colorrgb#1{\color{black}}%
      \def\colorgray#1{\color[gray]{#1}}%
      \expandafter\def\csname LTw\endcsname{\color{white}}%
      \expandafter\def\csname LTb\endcsname{\color{black}}%
      \expandafter\def\csname LTa\endcsname{\color{black}}%
      \expandafter\def\csname LT0\endcsname{\color{black}}%
      \expandafter\def\csname LT1\endcsname{\color{black}}%
      \expandafter\def\csname LT2\endcsname{\color{black}}%
      \expandafter\def\csname LT3\endcsname{\color{black}}%
      \expandafter\def\csname LT4\endcsname{\color{black}}%
      \expandafter\def\csname LT5\endcsname{\color{black}}%
      \expandafter\def\csname LT6\endcsname{\color{black}}%
      \expandafter\def\csname LT7\endcsname{\color{black}}%
      \expandafter\def\csname LT8\endcsname{\color{black}}%
    \fi
  \fi
    \setlength{\unitlength}{0.0500bp}%
    \ifx\gptboxheight\undefined%
      \newlength{\gptboxheight}%
      \newlength{\gptboxwidth}%
      \newsavebox{\gptboxtext}%
    \fi%
    \setlength{\fboxrule}{0.5pt}%
    \setlength{\fboxsep}{1pt}%
\begin{picture}(5760.00,6622.00)%
    \gplgaddtomacro\gplbacktext{%
      \csname LTb\endcsname
      \put(792,3973){\makebox(0,0)[r]{\strut{}0}}%
      \put(792,4459){\makebox(0,0)[r]{\strut{}$2 \cdot 10^{4}$}}%
      \put(792,4944){\makebox(0,0)[r]{\strut{}$4 \cdot 10^{4}$}}%
      \put(792,5430){\makebox(0,0)[r]{\strut{}$6 \cdot 10^{4}$}}%
      \put(792,5915){\makebox(0,0)[r]{\strut{}$8 \cdot 10^{4}$}}%
      \put(792,6401){\makebox(0,0)[r]{\strut{}$1 \cdot 10^{5}$}}%
      \put(1288,3753){\makebox(0,0){\strut{}}}%
      \put(1895,3753){\makebox(0,0){\strut{}}}%
      \put(2501,3753){\makebox(0,0){\strut{}}}%
      \put(3108,3753){\makebox(0,0){\strut{}}}%
      \put(3714,3753){\makebox(0,0){\strut{}}}%
      \put(4321,3753){\makebox(0,0){\strut{}}}%
      \put(4927,3753){\makebox(0,0){\strut{}}}%
      \put(5534,3753){\makebox(0,0){\strut{}}}%
    }%
    \gplgaddtomacro\gplfronttext{%
      \csname LTb\endcsname
      \put(-88,5187){\rotatebox{-270}{\makebox(0,0){\strut{} \scalebox{1.2}{$ {\rm d} \sigma / {\rm d} p_{\perp,\ell\ell} ~ [{\rm fb} / {\rm GeV}] $}}}}%
      \csname LTb\endcsname
      \put(4547,6173){\makebox(0,0)[r]{\strut{} \scalebox{1.2}{$\sigma_{\rm LO}^{\phantom{()}}+\sigma_{\rm NLO}^{(\alpha_s)}$}}}%
    }%
    \gplgaddtomacro\gplbacktext{%
      \csname LTb\endcsname
      \put(792,2325){\makebox(0,0)[r]{\strut{}0}}%
      \put(792,670){\makebox(0,0)[r]{\strut{}$-80$}}%
      \put(792,1084){\makebox(0,0)[r]{\strut{}$-60$}}%
      \put(792,1497){\makebox(0,0)[r]{\strut{}$-40$}}%
      \put(792,1911){\makebox(0,0)[r]{\strut{}$-20$}}%
      \put(792,2739){\makebox(0,0)[r]{\strut{}$20$}}%
      \put(792,3152){\makebox(0,0)[r]{\strut{}$40$}}%
      \put(792,3566){\makebox(0,0)[r]{\strut{}$60$}}%
      \put(1288,243){\makebox(0,0){\strut{}$10$}}%
      \put(1895,243){\makebox(0,0){\strut{}$20$}}%
      \put(2501,243){\makebox(0,0){\strut{}$30$}}%
      \put(3108,243){\makebox(0,0){\strut{}$40$}}%
      \put(3714,243){\makebox(0,0){\strut{}$50$}}%
      \put(4321,243){\makebox(0,0){\strut{}$60$}}%
      \put(4927,243){\makebox(0,0){\strut{}$70$}}%
      \put(5534,243){\makebox(0,0){\strut{}$80$}}%
    }%
    \gplgaddtomacro\gplfronttext{%
      \csname LTb\endcsname
      \put(176,2118){\rotatebox{-270}{\makebox(0,0){\strut{} \scalebox{1.2}{$ {\rm d} \Delta / {\rm d} p_{\perp,\ell\ell} ~ [\%] $}}}}%
      \put(3229,-87){\makebox(0,0){\strut{} \scalebox{1.2}{$ p_{\perp,\ell\ell} ~ [{\rm GeV}] $}}}%
      \csname LTb\endcsname
      \put(4547,1351){\makebox(0,0)[r]{\strut{}\scalebox{1.2}{$\Delta_{\alpha_s^2}$}}}%
      \csname LTb\endcsname
      \put(4547,1021){\makebox(0,0)[r]{\strut{}\scalebox{1.2}{$100 \times [\Delta_{\alpha_s \alpha}]_{P\otimes P}$}}}%
      \csname LTb\endcsname
      \put(4547,691){\makebox(0,0)[r]{\strut{}\scalebox{1.2}{$10\times [\Delta_{\alpha_s \alpha}]_{P\otimes D}$}}}%
    }%
    \gplbacktext
    \put(0,0){\includegraphics{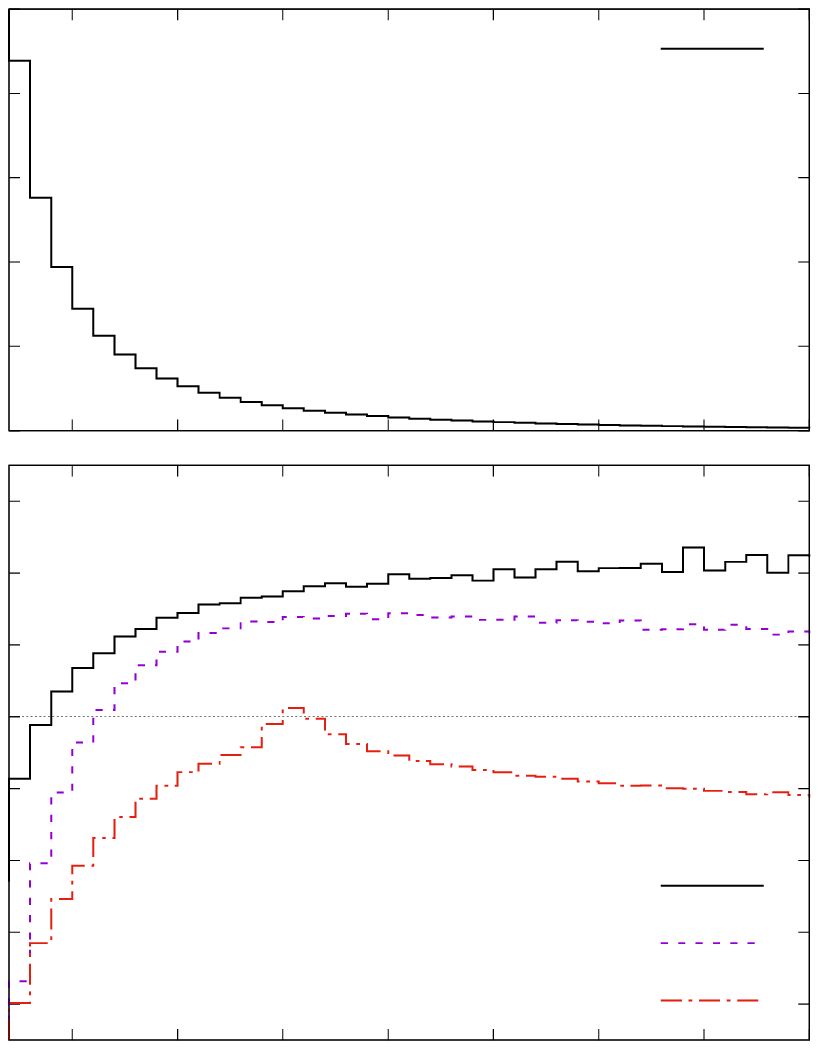}}%
    \gplfronttext
  \end{picture}%
\endgroup

%% file: figures/plots/DY_Histos_2108X_yll.tex
\begingroup
  \makeatletter
  \providecommand\color[2][]{%
    \GenericError{(gnuplot) \space\space\space\@spaces}{%
      Package color not loaded in conjunction with
      terminal option `colourtext'%
    }{See the gnuplot documentation for explanation.%
    }{Either use 'blacktext' in gnuplot or load the package
      color.sty in LaTeX.}%
    \renewcommand\color[2][]{}%
  }%
  \providecommand\includegraphics[2][]{%
    \GenericError{(gnuplot) \space\space\space\@spaces}{%
      Package graphicx or graphics not loaded%
    }{See the gnuplot documentation for explanation.%
    }{The gnuplot epslatex terminal needs graphicx.sty or graphics.sty.}%
    \renewcommand\includegraphics[2][]{}%
  }%
  \providecommand\rotatebox[2]{#2}%
  \@ifundefined{ifGPcolor}{%
    \newif\ifGPcolor
    \GPcolortrue
  }{}%
  \@ifundefined{ifGPblacktext}{%
    \newif\ifGPblacktext
    \GPblacktexttrue
  }{}%
  \let\gplgaddtomacro\g@addto@macro
  \gdef\gplbacktext{}%
  \gdef\gplfronttext{}%
  \makeatother
  \ifGPblacktext
    \def\colorrgb#1{}%
    \def\colorgray#1{}%
  \else
    \ifGPcolor
      \def\colorrgb#1{\color[rgb]{#1}}%
      \def\colorgray#1{\color[gray]{#1}}%
      \expandafter\def\csname LTw\endcsname{\color{white}}%
      \expandafter\def\csname LTb\endcsname{\color{black}}%
      \expandafter\def\csname LTa\endcsname{\color{black}}%
      \expandafter\def\csname LT0\endcsname{\color[rgb]{1,0,0}}%
      \expandafter\def\csname LT1\endcsname{\color[rgb]{0,1,0}}%
      \expandafter\def\csname LT2\endcsname{\color[rgb]{0,0,1}}%
      \expandafter\def\csname LT3\endcsname{\color[rgb]{1,0,1}}%
      \expandafter\def\csname LT4\endcsname{\color[rgb]{0,1,1}}%
      \expandafter\def\csname LT5\endcsname{\color[rgb]{1,1,0}}%
      \expandafter\def\csname LT6\endcsname{\color[rgb]{0,0,0}}%
      \expandafter\def\csname LT7\endcsname{\color[rgb]{1,0.3,0}}%
      \expandafter\def\csname LT8\endcsname{\color[rgb]{0.5,0.5,0.5}}%
    \else
      \def\colorrgb#1{\color{black}}%
      \def\colorgray#1{\color[gray]{#1}}%
      \expandafter\def\csname LTw\endcsname{\color{white}}%
      \expandafter\def\csname LTb\endcsname{\color{black}}%
      \expandafter\def\csname LTa\endcsname{\color{black}}%
      \expandafter\def\csname LT0\endcsname{\color{black}}%
      \expandafter\def\csname LT1\endcsname{\color{black}}%
      \expandafter\def\csname LT2\endcsname{\color{black}}%
      \expandafter\def\csname LT3\endcsname{\color{black}}%
      \expandafter\def\csname LT4\endcsname{\color{black}}%
      \expandafter\def\csname LT5\endcsname{\color{black}}%
      \expandafter\def\csname LT6\endcsname{\color{black}}%
      \expandafter\def\csname LT7\endcsname{\color{black}}%
      \expandafter\def\csname LT8\endcsname{\color{black}}%
    \fi
  \fi
    \setlength{\unitlength}{0.0500bp}%
    \ifx\gptboxheight\undefined%
      \newlength{\gptboxheight}%
      \newlength{\gptboxwidth}%
      \newsavebox{\gptboxtext}%
    \fi%
    \setlength{\fboxrule}{0.5pt}%
    \setlength{\fboxsep}{1pt}%
\begin{picture}(5760.00,6622.00)%
    \gplgaddtomacro\gplbacktext{%
      \csname LTb\endcsname
      \put(608,3753){\makebox(0,0){\strut{}}}%
      \put(1089,3753){\makebox(0,0){\strut{}}}%
      \put(1569,3753){\makebox(0,0){\strut{}}}%
      \put(2049,3753){\makebox(0,0){\strut{}}}%
      \put(2530,3753){\makebox(0,0){\strut{}}}%
      \put(3010,3753){\makebox(0,0){\strut{}}}%
      \put(3490,3753){\makebox(0,0){\strut{}}}%
      \put(3970,3753){\makebox(0,0){\strut{}}}%
      \put(4451,3753){\makebox(0,0){\strut{}}}%
      \put(4967,3973){\makebox(0,0)[l]{\strut{}0}}%
      \put(4967,4687){\makebox(0,0)[l]{\strut{}$1 \cdot 10^{5}$}}%
      \put(4967,5401){\makebox(0,0)[l]{\strut{}$2 \cdot 10^{5}$}}%
      \put(4967,6115){\makebox(0,0)[l]{\strut{}$3 \cdot 10^{5}$}}%
    }%
    \gplgaddtomacro\gplfronttext{%
      \csname LTb\endcsname
      \put(5891,5187){\rotatebox{-270}{\makebox(0,0){\strut{} \scalebox{1.2}{${\rm d} \sigma / {\rm d} y_{\ell\ell} ~ [{\rm fb}] $}}}}%
      \csname LTb\endcsname
      \put(3848,6173){\makebox(0,0)[r]{\strut{} \scalebox{1.2}{$\sigma_{\rm LO}^{\phantom{()}}+\sigma_{\rm NLO}^{(\alpha_s)}$}}}%
    }%
    \gplgaddtomacro\gplbacktext{%
      \csname LTb\endcsname
      \put(224,243){\makebox(0,0){\strut{}$-2.4$}}%
      \put(800,243){\makebox(0,0){\strut{}$-1.8$}}%
      \put(1377,243){\makebox(0,0){\strut{}$-1.2$}}%
      \put(1953,243){\makebox(0,0){\strut{}$-0.6$}}%
      \put(2530,243){\makebox(0,0){\strut{}$0$}}%
      \put(3106,243){\makebox(0,0){\strut{}$0.6$}}%
      \put(3682,243){\makebox(0,0){\strut{}$1.2$}}%
      \put(4259,243){\makebox(0,0){\strut{}$1.8$}}%
      \put(4835,243){\makebox(0,0){\strut{}$2.4$}}%
      \put(4967,613){\makebox(0,0)[l]{\strut{}$-8$}}%
      \put(4967,914){\makebox(0,0)[l]{\strut{}$-6$}}%
      \put(4967,1215){\makebox(0,0)[l]{\strut{}$-4$}}%
      \put(4967,1516){\makebox(0,0)[l]{\strut{}$-2$}}%
      \put(4967,1817){\makebox(0,0)[l]{\strut{}$0$}}%
      \put(4967,2118){\makebox(0,0)[l]{\strut{}$2$}}%
      \put(4967,2419){\makebox(0,0)[l]{\strut{}$4$}}%
      \put(4967,2720){\makebox(0,0)[l]{\strut{}$6$}}%
      \put(4967,3021){\makebox(0,0)[l]{\strut{}$8$}}%
      \put(4967,3322){\makebox(0,0)[l]{\strut{}$10$}}%
      \put(4967,3623){\makebox(0,0)[l]{\strut{}$12$}}%
    }%
    \gplgaddtomacro\gplfronttext{%
      \csname LTb\endcsname
      \put(5627,2118){\rotatebox{-270}{\makebox(0,0){\strut{} \scalebox{1.2}{${\rm d} \Delta / {\rm d} y_{\ell\ell} ~ [\%] $}}}}%
      \put(2529,-87){\makebox(0,0){\strut{} \scalebox{1.2}{$ y_{\ell\ell} $}}}%
      \csname LTb\endcsname
      \put(3848,3545){\makebox(0,0)[r]{\strut{} \scalebox{1.2}{$\Delta_{\alpha_s^2}$}}}%
      \csname LTb\endcsname
      \put(3848,3215){\makebox(0,0)[r]{\strut{} \scalebox{1.2}{$100 \times [\Delta_{\alpha_s \alpha}]_{P\otimes P}$}}}%
      \csname LTb\endcsname
      \put(3848,2885){\makebox(0,0)[r]{\strut{} \scalebox{1.2}{$10 \times [\Delta_{\alpha_s \alpha}]_{P\otimes D}$}}}%
    }%
    \gplbacktext
    \put(0,0){\includegraphics{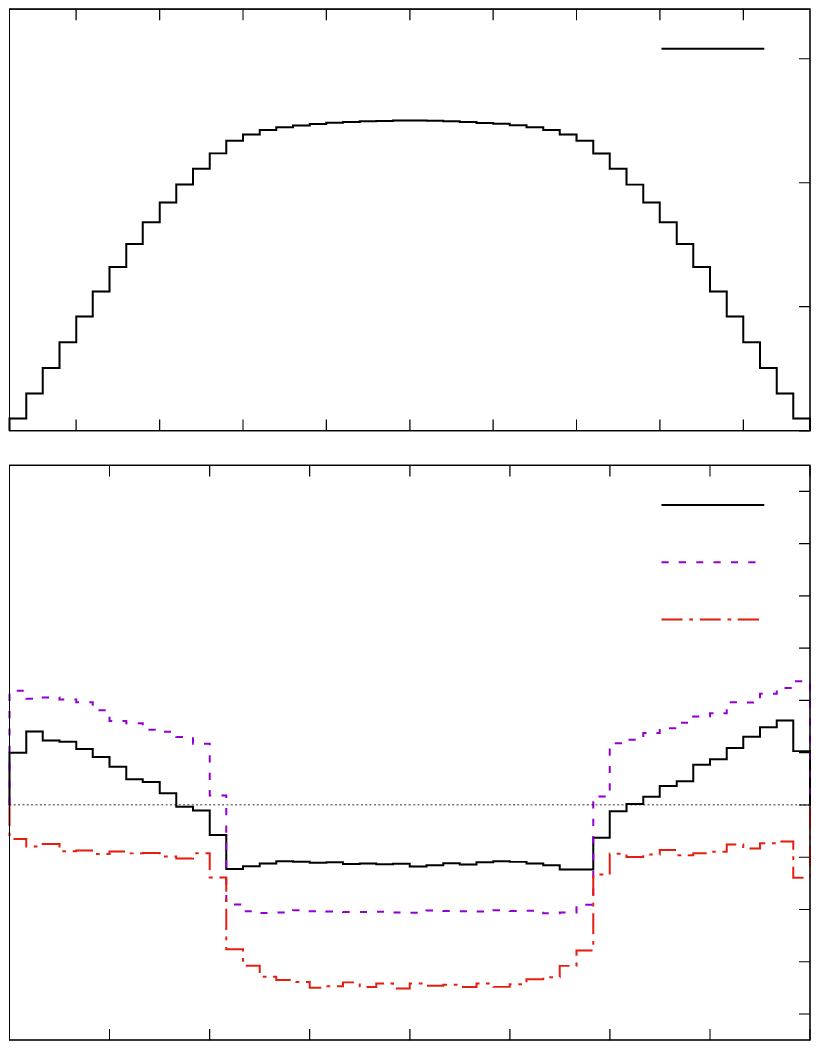}}%
    \gplfronttext
  \end{picture}%
\endgroup

%% file: figures/plots/DY_Histos_2108X_ptl1.tex
\begingroup
  \makeatletter
  \providecommand\color[2][]{%
    \GenericError{(gnuplot) \space\space\space\@spaces}{%
      Package color not loaded in conjunction with
      terminal option `colourtext'%
    }{See the gnuplot documentation for explanation.%
    }{Either use 'blacktext' in gnuplot or load the package
      color.sty in LaTeX.}%
    \renewcommand\color[2][]{}%
  }%
  \providecommand\includegraphics[2][]{%
    \GenericError{(gnuplot) \space\space\space\@spaces}{%
      Package graphicx or graphics not loaded%
    }{See the gnuplot documentation for explanation.%
    }{The gnuplot epslatex terminal needs graphicx.sty or graphics.sty.}%
    \renewcommand\includegraphics[2][]{}%
  }%
  \providecommand\rotatebox[2]{#2}%
  \@ifundefined{ifGPcolor}{%
    \newif\ifGPcolor
    \GPcolortrue
  }{}%
  \@ifundefined{ifGPblacktext}{%
    \newif\ifGPblacktext
    \GPblacktexttrue
  }{}%
  \let\gplgaddtomacro\g@addto@macro
  \gdef\gplbacktext{}%
  \gdef\gplfronttext{}%
  \makeatother
  \ifGPblacktext
    \def\colorrgb#1{}%
    \def\colorgray#1{}%
  \else
    \ifGPcolor
      \def\colorrgb#1{\color[rgb]{#1}}%
      \def\colorgray#1{\color[gray]{#1}}%
      \expandafter\def\csname LTw\endcsname{\color{white}}%
      \expandafter\def\csname LTb\endcsname{\color{black}}%
      \expandafter\def\csname LTa\endcsname{\color{black}}%
      \expandafter\def\csname LT0\endcsname{\color[rgb]{1,0,0}}%
      \expandafter\def\csname LT1\endcsname{\color[rgb]{0,1,0}}%
      \expandafter\def\csname LT2\endcsname{\color[rgb]{0,0,1}}%
      \expandafter\def\csname LT3\endcsname{\color[rgb]{1,0,1}}%
      \expandafter\def\csname LT4\endcsname{\color[rgb]{0,1,1}}%
      \expandafter\def\csname LT5\endcsname{\color[rgb]{1,1,0}}%
      \expandafter\def\csname LT6\endcsname{\color[rgb]{0,0,0}}%
      \expandafter\def\csname LT7\endcsname{\color[rgb]{1,0.3,0}}%
      \expandafter\def\csname LT8\endcsname{\color[rgb]{0.5,0.5,0.5}}%
    \else
      \def\colorrgb#1{\color{black}}%
      \def\colorgray#1{\color[gray]{#1}}%
      \expandafter\def\csname LTw\endcsname{\color{white}}%
      \expandafter\def\csname LTb\endcsname{\color{black}}%
      \expandafter\def\csname LTa\endcsname{\color{black}}%
      \expandafter\def\csname LT0\endcsname{\color{black}}%
      \expandafter\def\csname LT1\endcsname{\color{black}}%
      \expandafter\def\csname LT2\endcsname{\color{black}}%
      \expandafter\def\csname LT3\endcsname{\color{black}}%
      \expandafter\def\csname LT4\endcsname{\color{black}}%
      \expandafter\def\csname LT5\endcsname{\color{black}}%
      \expandafter\def\csname LT6\endcsname{\color{black}}%
      \expandafter\def\csname LT7\endcsname{\color{black}}%
      \expandafter\def\csname LT8\endcsname{\color{black}}%
    \fi
  \fi
    \setlength{\unitlength}{0.0500bp}%
    \ifx\gptboxheight\undefined%
      \newlength{\gptboxheight}%
      \newlength{\gptboxwidth}%
      \newsavebox{\gptboxtext}%
    \fi%
    \setlength{\fboxrule}{0.5pt}%
    \setlength{\fboxsep}{1pt}%
\begin{picture}(5760.00,6622.00)%
    \gplgaddtomacro\gplbacktext{%
      \csname LTb\endcsname
      \put(792,3973){\makebox(0,0)[r]{\strut{}0}}%
      \put(792,4259){\makebox(0,0)[r]{\strut{}$1 \cdot 10^{4}$}}%
      \put(792,4544){\makebox(0,0)[r]{\strut{}$2 \cdot 10^{4}$}}%
      \put(792,4830){\makebox(0,0)[r]{\strut{}$3 \cdot 10^{4}$}}%
      \put(792,5116){\makebox(0,0)[r]{\strut{}$4 \cdot 10^{4}$}}%
      \put(792,5401){\makebox(0,0)[r]{\strut{}$5 \cdot 10^{4}$}}%
      \put(792,5687){\makebox(0,0)[r]{\strut{}$6 \cdot 10^{4}$}}%
      \put(792,5973){\makebox(0,0)[r]{\strut{}$7 \cdot 10^{4}$}}%
      \put(792,6258){\makebox(0,0)[r]{\strut{}$8 \cdot 10^{4}$}}%
      \put(1288,3753){\makebox(0,0){\strut{}}}%
      \put(1895,3753){\makebox(0,0){\strut{}}}%
      \put(2501,3753){\makebox(0,0){\strut{}}}%
      \put(3108,3753){\makebox(0,0){\strut{}}}%
      \put(3714,3753){\makebox(0,0){\strut{}}}%
      \put(4321,3753){\makebox(0,0){\strut{}}}%
      \put(4927,3753){\makebox(0,0){\strut{}}}%
      \put(5534,3753){\makebox(0,0){\strut{}}}%
    }%
    \gplgaddtomacro\gplfronttext{%
      \csname LTb\endcsname
      \put(-88,5187){\rotatebox{-270}{\makebox(0,0){\strut{} \scalebox{1.2}{${\rm d} \sigma / {\rm d} p_{\perp,\ell_1} ~ [{\rm fb} / {\rm GeV}] $}}}}%
      \csname LTb\endcsname
      \put(4547,6173){\makebox(0,0)[r]{\strut{} \scalebox{1.2}{$\sigma_{\rm LO}^{\phantom{()}}+\sigma_{\rm NLO}^{(\alpha_s)}$}}}%
    }%
    \gplgaddtomacro\gplbacktext{%
      \csname LTb\endcsname
      \put(792,610){\makebox(0,0)[r]{\strut{}$-80$}}%
      \put(792,904){\makebox(0,0)[r]{\strut{}$-60$}}%
      \put(792,1199){\makebox(0,0)[r]{\strut{}$-40$}}%
      \put(792,1493){\makebox(0,0)[r]{\strut{}$-20$}}%
      \put(792,1787){\makebox(0,0)[r]{\strut{}$0$}}%
      \put(792,2081){\makebox(0,0)[r]{\strut{}$20$}}%
      \put(792,2375){\makebox(0,0)[r]{\strut{}$40$}}%
      \put(792,2670){\makebox(0,0)[r]{\strut{}$60$}}%
      \put(792,2964){\makebox(0,0)[r]{\strut{}$80$}}%
      \put(792,3258){\makebox(0,0)[r]{\strut{}$100$}}%
      \put(792,3552){\makebox(0,0)[r]{\strut{}$120$}}%
      \put(1288,243){\makebox(0,0){\strut{}$30$}}%
      \put(1895,243){\makebox(0,0){\strut{}$40$}}%
      \put(2501,243){\makebox(0,0){\strut{}$50$}}%
      \put(3108,243){\makebox(0,0){\strut{}$60$}}%
      \put(3714,243){\makebox(0,0){\strut{}$70$}}%
      \put(4321,243){\makebox(0,0){\strut{}$80$}}%
      \put(4927,243){\makebox(0,0){\strut{}$90$}}%
      \put(5534,243){\makebox(0,0){\strut{}$100$}}%
    }%
    \gplgaddtomacro\gplfronttext{%
      \csname LTb\endcsname
      \put(176,2118){\rotatebox{-270}{\makebox(0,0){\strut{} \scalebox{1.2}{${\rm d} \Delta / {\rm d} p_{\perp,\ell_1} ~ [\%] $}}}}%
      \put(3229,-87){\makebox(0,0){\strut{} \scalebox{1.2}{$ p_{\perp,\ell_1} ~ [{\rm GeV}] $}}}%
      \csname LTb\endcsname
      \put(4547,3545){\makebox(0,0)[r]{\strut{} \scalebox{1.2}{$\Delta_{\alpha_s^2}$}}}%
      \csname LTb\endcsname
      \put(4547,3215){\makebox(0,0)[r]{\strut{} \scalebox{1.2}{$100 \times [\Delta_{\alpha_s \alpha}]_{P\otimes P}$}}}%
      \csname LTb\endcsname
      \put(4547,2885){\makebox(0,0)[r]{\strut{} \scalebox{1.2}{$ 10 \times [\Delta_{\alpha_s \alpha}]_{P\otimes D}$}}}%
    }%
    \gplbacktext
    \put(0,0){\includegraphics{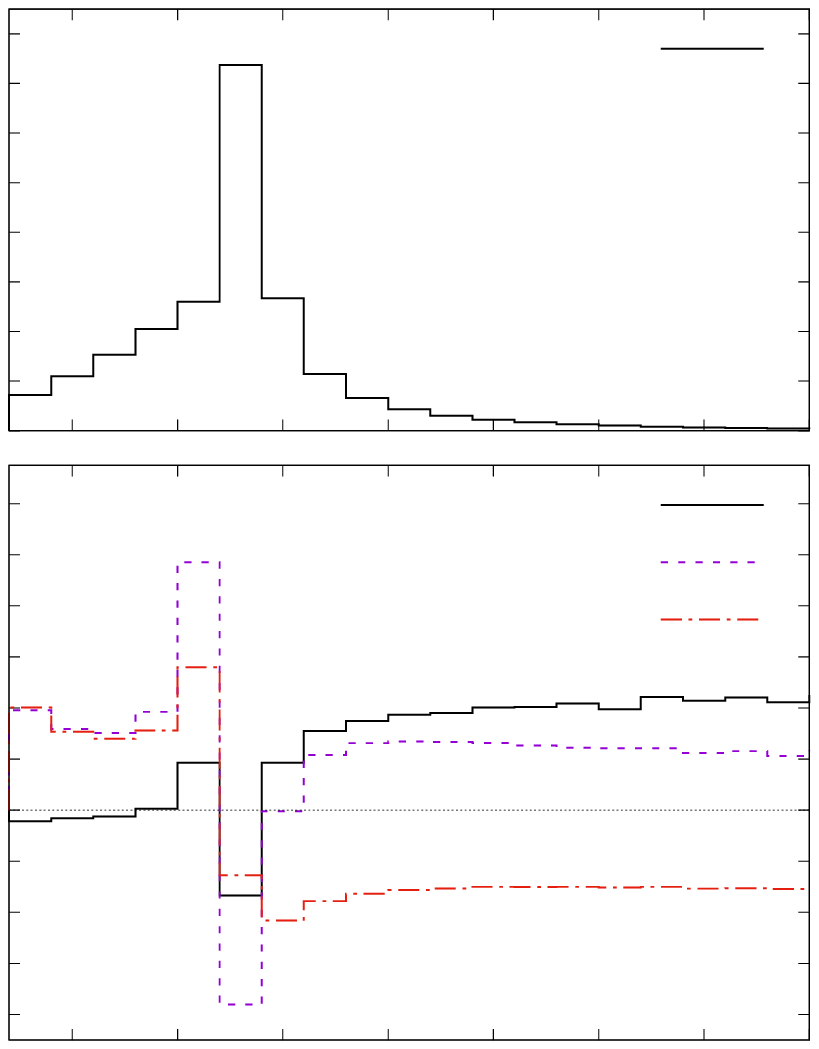}}%
    \gplfronttext
  \end{picture}%
\endgroup

%% file: figures/plots/DY_Histos_2108X_ptl2.tex
\begingroup
  \makeatletter
  \providecommand\color[2][]{%
    \GenericError{(gnuplot) \space\space\space\@spaces}{%
      Package color not loaded in conjunction with
      terminal option `colourtext'%
    }{See the gnuplot documentation for explanation.%
    }{Either use 'blacktext' in gnuplot or load the package
      color.sty in LaTeX.}%
    \renewcommand\color[2][]{}%
  }%
  \providecommand\includegraphics[2][]{%
    \GenericError{(gnuplot) \space\space\space\@spaces}{%
      Package graphicx or graphics not loaded%
    }{See the gnuplot documentation for explanation.%
    }{The gnuplot epslatex terminal needs graphicx.sty or graphics.sty.}%
    \renewcommand\includegraphics[2][]{}%
  }%
  \providecommand\rotatebox[2]{#2}%
  \@ifundefined{ifGPcolor}{%
    \newif\ifGPcolor
    \GPcolortrue
  }{}%
  \@ifundefined{ifGPblacktext}{%
    \newif\ifGPblacktext
    \GPblacktexttrue
  }{}%
  \let\gplgaddtomacro\g@addto@macro
  \gdef\gplbacktext{}%
  \gdef\gplfronttext{}%
  \makeatother
  \ifGPblacktext
    \def\colorrgb#1{}%
    \def\colorgray#1{}%
  \else
    \ifGPcolor
      \def\colorrgb#1{\color[rgb]{#1}}%
      \def\colorgray#1{\color[gray]{#1}}%
      \expandafter\def\csname LTw\endcsname{\color{white}}%
      \expandafter\def\csname LTb\endcsname{\color{black}}%
      \expandafter\def\csname LTa\endcsname{\color{black}}%
      \expandafter\def\csname LT0\endcsname{\color[rgb]{1,0,0}}%
      \expandafter\def\csname LT1\endcsname{\color[rgb]{0,1,0}}%
      \expandafter\def\csname LT2\endcsname{\color[rgb]{0,0,1}}%
      \expandafter\def\csname LT3\endcsname{\color[rgb]{1,0,1}}%
      \expandafter\def\csname LT4\endcsname{\color[rgb]{0,1,1}}%
      \expandafter\def\csname LT5\endcsname{\color[rgb]{1,1,0}}%
      \expandafter\def\csname LT6\endcsname{\color[rgb]{0,0,0}}%
      \expandafter\def\csname LT7\endcsname{\color[rgb]{1,0.3,0}}%
      \expandafter\def\csname LT8\endcsname{\color[rgb]{0.5,0.5,0.5}}%
    \else
      \def\colorrgb#1{\color{black}}%
      \def\colorgray#1{\color[gray]{#1}}%
      \expandafter\def\csname LTw\endcsname{\color{white}}%
      \expandafter\def\csname LTb\endcsname{\color{black}}%
      \expandafter\def\csname LTa\endcsname{\color{black}}%
      \expandafter\def\csname LT0\endcsname{\color{black}}%
      \expandafter\def\csname LT1\endcsname{\color{black}}%
      \expandafter\def\csname LT2\endcsname{\color{black}}%
      \expandafter\def\csname LT3\endcsname{\color{black}}%
      \expandafter\def\csname LT4\endcsname{\color{black}}%
      \expandafter\def\csname LT5\endcsname{\color{black}}%
      \expandafter\def\csname LT6\endcsname{\color{black}}%
      \expandafter\def\csname LT7\endcsname{\color{black}}%
      \expandafter\def\csname LT8\endcsname{\color{black}}%
    \fi
  \fi
    \setlength{\unitlength}{0.0500bp}%
    \ifx\gptboxheight\undefined%
      \newlength{\gptboxheight}%
      \newlength{\gptboxwidth}%
      \newsavebox{\gptboxtext}%
    \fi%
    \setlength{\fboxrule}{0.5pt}%
    \setlength{\fboxsep}{1pt}%
\begin{picture}(5760.00,6622.00)%
    \gplgaddtomacro\gplbacktext{%
      \csname LTb\endcsname
      \put(92,3973){\makebox(0,0)[r]{\strut{}}}%
      \put(92,4414){\makebox(0,0)[r]{\strut{}}}%
      \put(92,4856){\makebox(0,0)[r]{\strut{}}}%
      \put(92,5297){\makebox(0,0)[r]{\strut{}}}%
      \put(92,5739){\makebox(0,0)[r]{\strut{}}}%
      \put(92,6180){\makebox(0,0)[r]{\strut{}}}%
      \put(444,3753){\makebox(0,0){\strut{}}}%
      \put(993,3753){\makebox(0,0){\strut{}}}%
      \put(1541,3753){\makebox(0,0){\strut{}}}%
      \put(2090,3753){\makebox(0,0){\strut{}}}%
      \put(2639,3753){\makebox(0,0){\strut{}}}%
      \put(3188,3753){\makebox(0,0){\strut{}}}%
      \put(3737,3753){\makebox(0,0){\strut{}}}%
      \put(4286,3753){\makebox(0,0){\strut{}}}%
      \put(4835,3753){\makebox(0,0){\strut{}}}%
      \put(4967,3973){\makebox(0,0)[l]{\strut{}0}}%
      \put(4967,4414){\makebox(0,0)[l]{\strut{}$1 \cdot 10^{4}$}}%
      \put(4967,4856){\makebox(0,0)[l]{\strut{}$2 \cdot 10^{4}$}}%
      \put(4967,5297){\makebox(0,0)[l]{\strut{}$3 \cdot 10^{4}$}}%
      \put(4967,5739){\makebox(0,0)[l]{\strut{}$4 \cdot 10^{4}$}}%
      \put(4967,6180){\makebox(0,0)[l]{\strut{}$5 \cdot 10^{4}$}}%
    }%
    \gplgaddtomacro\gplfronttext{%
      \csname LTb\endcsname
      \put(5891,5187){\rotatebox{-270}{\makebox(0,0){\strut{} \scalebox{1.2}{${\rm d} \sigma / {\rm d} p_{\perp,\ell_2} ~ [{\rm fb} / {\rm GeV}] $}}}}%
      \csname LTb\endcsname
      \put(3848,6173){\makebox(0,0)[r]{\strut{} \scalebox{1.2}{$\sigma_{\rm LO}^{\phantom{()}}+\sigma_{\rm NLO}^{(\alpha_s)}$}}}%
    }%
    \gplgaddtomacro\gplbacktext{%
      \csname LTb\endcsname
      \put(92,670){\makebox(0,0)[r]{\strut{}}}%
      \put(92,1084){\makebox(0,0)[r]{\strut{}}}%
      \put(92,1497){\makebox(0,0)[r]{\strut{}}}%
      \put(92,1911){\makebox(0,0)[r]{\strut{}}}%
      \put(92,2325){\makebox(0,0)[r]{\strut{}}}%
      \put(92,2739){\makebox(0,0)[r]{\strut{}}}%
      \put(92,3152){\makebox(0,0)[r]{\strut{}}}%
      \put(92,3566){\makebox(0,0)[r]{\strut{}}}%
      \put(444,243){\makebox(0,0){\strut{}$20$}}%
      \put(993,243){\makebox(0,0){\strut{}$30$}}%
      \put(1541,243){\makebox(0,0){\strut{}$40$}}%
      \put(2090,243){\makebox(0,0){\strut{}$50$}}%
      \put(2639,243){\makebox(0,0){\strut{}$60$}}%
      \put(3188,243){\makebox(0,0){\strut{}$70$}}%
      \put(3737,243){\makebox(0,0){\strut{}$80$}}%
      \put(4286,243){\makebox(0,0){\strut{}$90$}}%
      \put(4835,243){\makebox(0,0){\strut{}$100$}}%
      \put(4967,670){\makebox(0,0)[l]{\strut{}$-80$}}%
      \put(4967,1084){\makebox(0,0)[l]{\strut{}$-60$}}%
      \put(4967,1497){\makebox(0,0)[l]{\strut{}$-40$}}%
      \put(4967,1911){\makebox(0,0)[l]{\strut{}$-20$}}%
      \put(4967,2325){\makebox(0,0)[l]{\strut{}$0$}}%
      \put(4967,2739){\makebox(0,0)[l]{\strut{}$20$}}%
      \put(4967,3152){\makebox(0,0)[l]{\strut{}$40$}}%
      \put(4967,3566){\makebox(0,0)[l]{\strut{}$60$}}%
    }%
    \gplgaddtomacro\gplfronttext{%
      \csname LTb\endcsname
      \put(5627,2118){\rotatebox{-270}{\makebox(0,0){\strut{} \scalebox{1.2}{${\rm d} \Delta / {\rm d} p_{\perp,\ell_2} ~ [\%] $}}}}%
      \put(2529,-87){\makebox(0,0){\strut{} \scalebox{1.2}{$ p_{\perp,\ell_2} ~ [{\rm GeV}] $}}}%
      \csname LTb\endcsname
      \put(3848,1351){\makebox(0,0)[r]{\strut{} \scalebox{1.2}{$\Delta_{\alpha_s^2}$}}}%
      \csname LTb\endcsname
      \put(3848,1021){\makebox(0,0)[r]{\strut{} \scalebox{1.2}{$100 \times [\Delta_{\alpha_s \alpha}]_{P\otimes P}$}}}%
      \csname LTb\endcsname
      \put(3848,691){\makebox(0,0)[r]{\strut{} \scalebox{1.2}{$ 10 \times [\Delta_{\alpha_s \alpha}]_{P\otimes D}$}}}%
    }%
    \gplbacktext
    \put(0,0){\includegraphics{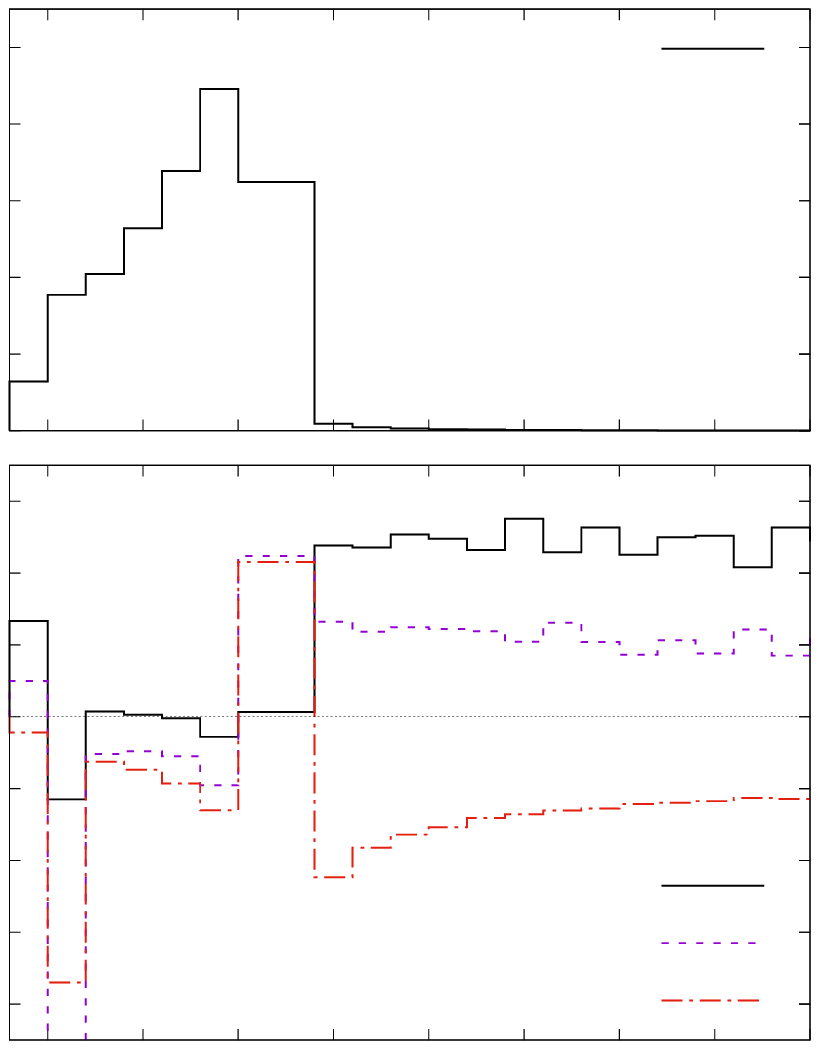}}%
    \gplfronttext
  \end{picture}%
\endgroup

%% file: figures/plots/DY_Histos_2108X_costh2.tex
\begingroup
  \makeatletter
  \providecommand\color[2][]{%
    \GenericError{(gnuplot) \space\space\space\@spaces}{%
      Package color not loaded in conjunction with
      terminal option `colourtext'%
    }{See the gnuplot documentation for explanation.%
    }{Either use 'blacktext' in gnuplot or load the package
      color.sty in LaTeX.}%
    \renewcommand\color[2][]{}%
  }%
  \providecommand\includegraphics[2][]{%
    \GenericError{(gnuplot) \space\space\space\@spaces}{%
      Package graphicx or graphics not loaded%
    }{See the gnuplot documentation for explanation.%
    }{The gnuplot epslatex terminal needs graphicx.sty or graphics.sty.}%
    \renewcommand\includegraphics[2][]{}%
  }%
  \providecommand\rotatebox[2]{#2}%
  \@ifundefined{ifGPcolor}{%
    \newif\ifGPcolor
    \GPcolortrue
  }{}%
  \@ifundefined{ifGPblacktext}{%
    \newif\ifGPblacktext
    \GPblacktexttrue
  }{}%
  \let\gplgaddtomacro\g@addto@macro
  \gdef\gplbacktext{}%
  \gdef\gplfronttext{}%
  \makeatother
  \ifGPblacktext
    \def\colorrgb#1{}%
    \def\colorgray#1{}%
  \else
    \ifGPcolor
      \def\colorrgb#1{\color[rgb]{#1}}%
      \def\colorgray#1{\color[gray]{#1}}%
      \expandafter\def\csname LTw\endcsname{\color{white}}%
      \expandafter\def\csname LTb\endcsname{\color{black}}%
      \expandafter\def\csname LTa\endcsname{\color{black}}%
      \expandafter\def\csname LT0\endcsname{\color[rgb]{1,0,0}}%
      \expandafter\def\csname LT1\endcsname{\color[rgb]{0,1,0}}%
      \expandafter\def\csname LT2\endcsname{\color[rgb]{0,0,1}}%
      \expandafter\def\csname LT3\endcsname{\color[rgb]{1,0,1}}%
      \expandafter\def\csname LT4\endcsname{\color[rgb]{0,1,1}}%
      \expandafter\def\csname LT5\endcsname{\color[rgb]{1,1,0}}%
      \expandafter\def\csname LT6\endcsname{\color[rgb]{0,0,0}}%
      \expandafter\def\csname LT7\endcsname{\color[rgb]{1,0.3,0}}%
      \expandafter\def\csname LT8\endcsname{\color[rgb]{0.5,0.5,0.5}}%
    \else
      \def\colorrgb#1{\color{black}}%
      \def\colorgray#1{\color[gray]{#1}}%
      \expandafter\def\csname LTw\endcsname{\color{white}}%
      \expandafter\def\csname LTb\endcsname{\color{black}}%
      \expandafter\def\csname LTa\endcsname{\color{black}}%
      \expandafter\def\csname LT0\endcsname{\color{black}}%
      \expandafter\def\csname LT1\endcsname{\color{black}}%
      \expandafter\def\csname LT2\endcsname{\color{black}}%
      \expandafter\def\csname LT3\endcsname{\color{black}}%
      \expandafter\def\csname LT4\endcsname{\color{black}}%
      \expandafter\def\csname LT5\endcsname{\color{black}}%
      \expandafter\def\csname LT6\endcsname{\color{black}}%
      \expandafter\def\csname LT7\endcsname{\color{black}}%
      \expandafter\def\csname LT8\endcsname{\color{black}}%
    \fi
  \fi
    \setlength{\unitlength}{0.0500bp}%
    \ifx\gptboxheight\undefined%
      \newlength{\gptboxheight}%
      \newlength{\gptboxwidth}%
      \newsavebox{\gptboxtext}%
    \fi%
    \setlength{\fboxrule}{0.5pt}%
    \setlength{\fboxsep}{1pt}%
\begin{picture}(5760.00,6622.00)%
    \gplgaddtomacro\gplbacktext{%
      \csname LTb\endcsname
      \put(792,3973){\makebox(0,0)[r]{\strut{}0}}%
      \put(792,4479){\makebox(0,0)[r]{\strut{}$5.0 \cdot 10^{4}$}}%
      \put(792,4985){\makebox(0,0)[r]{\strut{}$1.0 \cdot 10^{5}$}}%
      \put(792,5491){\makebox(0,0)[r]{\strut{}$1.5 \cdot 10^{5}$}}%
      \put(792,5996){\makebox(0,0)[r]{\strut{}$2.0 \cdot 10^{5}$}}%
      \put(924,3753){\makebox(0,0){\strut{}}}%
      \put(2077,3753){\makebox(0,0){\strut{}}}%
      \put(3229,3753){\makebox(0,0){\strut{}}}%
      \put(4382,3753){\makebox(0,0){\strut{}}}%
      \put(5534,3753){\makebox(0,0){\strut{}}}%
    }%
    \gplgaddtomacro\gplfronttext{%
      \csname LTb\endcsname
      \put(-220,5187){\rotatebox{-270}{\makebox(0,0){\strut{} \scalebox{1.2}{${\rm d} \sigma / {\rm d} \cos \theta^{*} ~  [{\rm fb}] $}}}}%
      \csname LTb\endcsname
      \put(4547,6173){\makebox(0,0)[r]{\strut{} \scalebox{1.2}{$\sigma_{\rm LO}^{\phantom{()}}+\sigma_{\rm NLO}^{(\alpha_s)}$}}}%
    }%
    \gplgaddtomacro\gplbacktext{%
      \csname LTb\endcsname
      \put(792,739){\makebox(0,0)[r]{\strut{}$-3$}}%
      \put(792,1291){\makebox(0,0)[r]{\strut{}$-1$}}%
      \put(792,1842){\makebox(0,0)[r]{\strut{}$1$}}%
      \put(792,2394){\makebox(0,0)[r]{\strut{}$3$}}%
      \put(792,2946){\makebox(0,0)[r]{\strut{}$5$}}%
      \put(792,3497){\makebox(0,0)[r]{\strut{}$7$}}%
      \put(924,243){\makebox(0,0){\strut{}$-1$}}%
      \put(2077,243){\makebox(0,0){\strut{}$-0.5$}}%
      \put(3229,243){\makebox(0,0){\strut{}$0$}}%
      \put(4382,243){\makebox(0,0){\strut{}$0.5$}}%
      \put(5534,243){\makebox(0,0){\strut{}$1$}}%
    }%
    \gplgaddtomacro\gplfronttext{%
      \csname LTb\endcsname
      \put(308,2118){\rotatebox{-270}{\makebox(0,0){\strut{} \scalebox{1.2}{${\rm d} \Delta / {\rm d} \cos \theta^{*} ~ [\%] $}}}}%
      \put(3229,-87){\makebox(0,0){\strut{} \scalebox{1.2}{$ \cos \theta^{*} $}}}%
      \csname LTb\endcsname
      \put(4547,3545){\makebox(0,0)[r]{\strut{} \scalebox{1.2}{$\Delta_{\alpha_s^2}$}}}%
      \csname LTb\endcsname
      \put(4547,3215){\makebox(0,0)[r]{\strut{} \scalebox{1.2}{$100 \times [\Delta_{\alpha_s \alpha}]_{P\otimes P}$}}}%
      \csname LTb\endcsname
      \put(4547,2885){\makebox(0,0)[r]{\strut{} \scalebox{1.2}{$10\times [\Delta_{\alpha_s \alpha}]_{P\otimes D}$}}}%
    }%
    \gplbacktext
    \put(0,0){\includegraphics{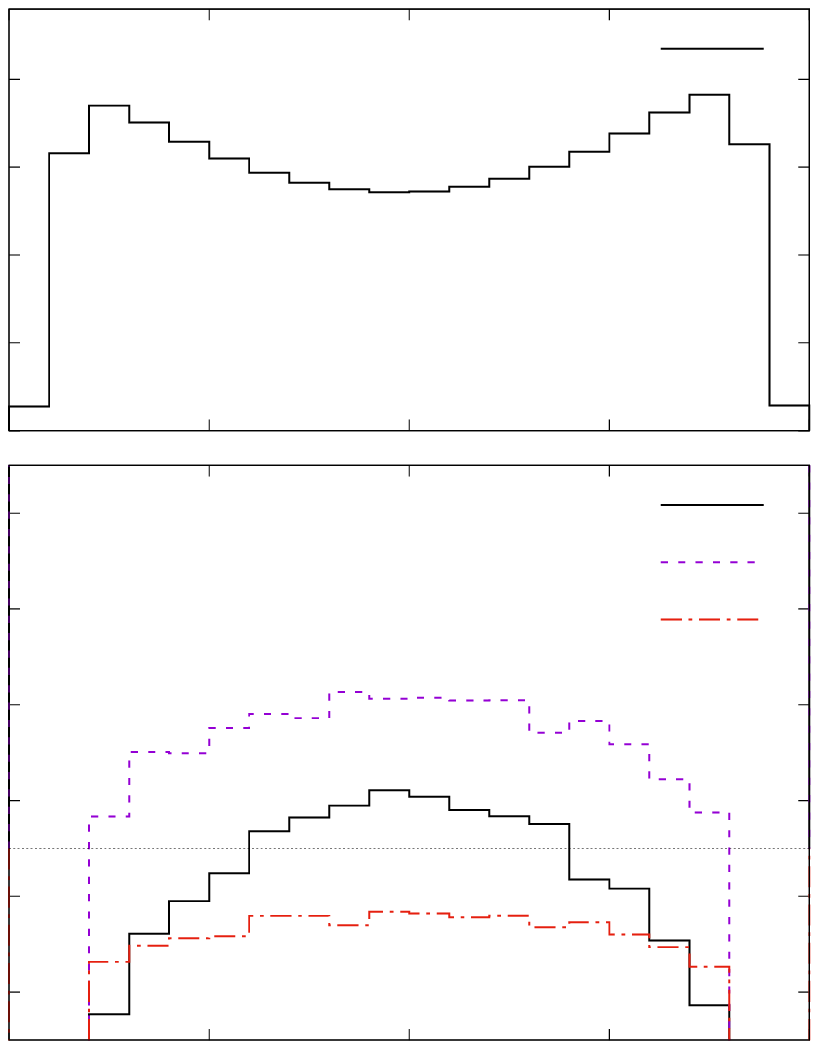}}%
    \gplfronttext
  \end{picture}%
\endgroup

%% file: figures/plots/DY_Histos_2108X_costh4.tex
\begingroup
  \makeatletter
  \providecommand\color[2][]{%
    \GenericError{(gnuplot) \space\space\space\@spaces}{%
      Package color not loaded in conjunction with
      terminal option `colourtext'%
    }{See the gnuplot documentation for explanation.%
    }{Either use 'blacktext' in gnuplot or load the package
      color.sty in LaTeX.}%
    \renewcommand\color[2][]{}%
  }%
  \providecommand\includegraphics[2][]{%
    \GenericError{(gnuplot) \space\space\space\@spaces}{%
      Package graphicx or graphics not loaded%
    }{See the gnuplot documentation for explanation.%
    }{The gnuplot epslatex terminal needs graphicx.sty or graphics.sty.}%
    \renewcommand\includegraphics[2][]{}%
  }%
  \providecommand\rotatebox[2]{#2}%
  \@ifundefined{ifGPcolor}{%
    \newif\ifGPcolor
    \GPcolortrue
  }{}%
  \@ifundefined{ifGPblacktext}{%
    \newif\ifGPblacktext
    \GPblacktexttrue
  }{}%
  \let\gplgaddtomacro\g@addto@macro
  \gdef\gplbacktext{}%
  \gdef\gplfronttext{}%
  \makeatother
  \ifGPblacktext
    \def\colorrgb#1{}%
    \def\colorgray#1{}%
  \else
    \ifGPcolor
      \def\colorrgb#1{\color[rgb]{#1}}%
      \def\colorgray#1{\color[gray]{#1}}%
      \expandafter\def\csname LTw\endcsname{\color{white}}%
      \expandafter\def\csname LTb\endcsname{\color{black}}%
      \expandafter\def\csname LTa\endcsname{\color{black}}%
      \expandafter\def\csname LT0\endcsname{\color[rgb]{1,0,0}}%
      \expandafter\def\csname LT1\endcsname{\color[rgb]{0,1,0}}%
      \expandafter\def\csname LT2\endcsname{\color[rgb]{0,0,1}}%
      \expandafter\def\csname LT3\endcsname{\color[rgb]{1,0,1}}%
      \expandafter\def\csname LT4\endcsname{\color[rgb]{0,1,1}}%
      \expandafter\def\csname LT5\endcsname{\color[rgb]{1,1,0}}%
      \expandafter\def\csname LT6\endcsname{\color[rgb]{0,0,0}}%
      \expandafter\def\csname LT7\endcsname{\color[rgb]{1,0.3,0}}%
      \expandafter\def\csname LT8\endcsname{\color[rgb]{0.5,0.5,0.5}}%
    \else
      \def\colorrgb#1{\color{black}}%
      \def\colorgray#1{\color[gray]{#1}}%
      \expandafter\def\csname LTw\endcsname{\color{white}}%
      \expandafter\def\csname LTb\endcsname{\color{black}}%
      \expandafter\def\csname LTa\endcsname{\color{black}}%
      \expandafter\def\csname LT0\endcsname{\color{black}}%
      \expandafter\def\csname LT1\endcsname{\color{black}}%
      \expandafter\def\csname LT2\endcsname{\color{black}}%
      \expandafter\def\csname LT3\endcsname{\color{black}}%
      \expandafter\def\csname LT4\endcsname{\color{black}}%
      \expandafter\def\csname LT5\endcsname{\color{black}}%
      \expandafter\def\csname LT6\endcsname{\color{black}}%
      \expandafter\def\csname LT7\endcsname{\color{black}}%
      \expandafter\def\csname LT8\endcsname{\color{black}}%
    \fi
  \fi
    \setlength{\unitlength}{0.0500bp}%
    \ifx\gptboxheight\undefined%
      \newlength{\gptboxheight}%
      \newlength{\gptboxwidth}%
      \newsavebox{\gptboxtext}%
    \fi%
    \setlength{\fboxrule}{0.5pt}%
    \setlength{\fboxsep}{1pt}%
\begin{picture}(5760.00,6622.00)%
    \gplgaddtomacro\gplbacktext{%
      \csname LTb\endcsname
      \put(224,3753){\makebox(0,0){\strut{}}}%
      \put(1377,3753){\makebox(0,0){\strut{}}}%
      \put(2530,3753){\makebox(0,0){\strut{}}}%
      \put(3682,3753){\makebox(0,0){\strut{}}}%
      \put(4835,3753){\makebox(0,0){\strut{}}}%
      \put(4967,3973){\makebox(0,0)[l]{\strut{}0}}%
      \put(4967,4297){\makebox(0,0)[l]{\strut{}$2.0 \cdot 10^{4}$}}%
      \put(4967,4620){\makebox(0,0)[l]{\strut{}$4.0 \cdot 10^{4}$}}%
      \put(4967,4944){\makebox(0,0)[l]{\strut{}$6.0 \cdot 10^{4}$}}%
      \put(4967,5268){\makebox(0,0)[l]{\strut{}$8.0 \cdot 10^{4}$}}%
      \put(4967,5592){\makebox(0,0)[l]{\strut{}$1.0 \cdot 10^{5}$}}%
      \put(4967,5915){\makebox(0,0)[l]{\strut{}$1.2 \cdot 10^{5}$}}%
      \put(4967,6239){\makebox(0,0)[l]{\strut{}$1.4 \cdot 10^{5}$}}%
    }%
    \gplgaddtomacro\gplfronttext{%
      \csname LTb\endcsname
      \put(6023,5187){\rotatebox{-270}{\makebox(0,0){\strut{} \scalebox{1.2}{${\rm d} \sigma / {\rm d} \cos \theta^{*} ~ [{\rm fb}] $}}}}%
      \csname LTb\endcsname
      \put(3848,6173){\makebox(0,0)[r]{\strut{} \scalebox{1.2}{$\sigma_{\rm LO}^{\phantom{()}}+\sigma_{\rm NLO}^{(\alpha_s)}$}}}%
    }%
    \gplgaddtomacro\gplbacktext{%
      \csname LTb\endcsname
      \put(224,243){\makebox(0,0){\strut{}$-0.5$}}%
      \put(1377,243){\makebox(0,0){\strut{}$-0.25$}}%
      \put(2530,243){\makebox(0,0){\strut{}$0$}}%
      \put(3682,243){\makebox(0,0){\strut{}$0.25$}}%
      \put(4835,243){\makebox(0,0){\strut{}$0.5$}}%
      \put(4967,699){\makebox(0,0)[l]{\strut{}$-3$}}%
      \put(4967,1172){\makebox(0,0)[l]{\strut{}$-1$}}%
      \put(4967,1645){\makebox(0,0)[l]{\strut{}$1$}}%
      \put(4967,2118){\makebox(0,0)[l]{\strut{}$3$}}%
      \put(4967,2591){\makebox(0,0)[l]{\strut{}$5$}}%
      \put(4967,3064){\makebox(0,0)[l]{\strut{}$7$}}%
      \put(4967,3537){\makebox(0,0)[l]{\strut{}$9$}}%
    }%
    \gplgaddtomacro\gplfronttext{%
      \csname LTb\endcsname
      \put(5495,2118){\rotatebox{-270}{\makebox(0,0){\strut{} \scalebox{1.2}{${\rm d} \Delta / {\rm d} \cos \theta^{*} ~ [\%] $}}}}%
      \put(2529,-87){\makebox(0,0){\strut{} \scalebox{1.2}{$ \cos \theta^{*} $}}}%
      \csname LTb\endcsname
      \put(3848,3545){\makebox(0,0)[r]{\strut{} \scalebox{1.2}{$\Delta_{\alpha_s^2}$}}}%
      \csname LTb\endcsname
      \put(3848,3215){\makebox(0,0)[r]{\strut{} \scalebox{1.2}{$100 \times [\Delta_{\alpha_s \alpha}]_{P\otimes P}$}}}%
      \csname LTb\endcsname
      \put(3848,2885){\makebox(0,0)[r]{\strut{} \scalebox{1.2}{$10 \times [\Delta_{\alpha_s \alpha}]_{P\otimes D}$}}}%
    }%
    \gplbacktext
    \put(0,0){\includegraphics{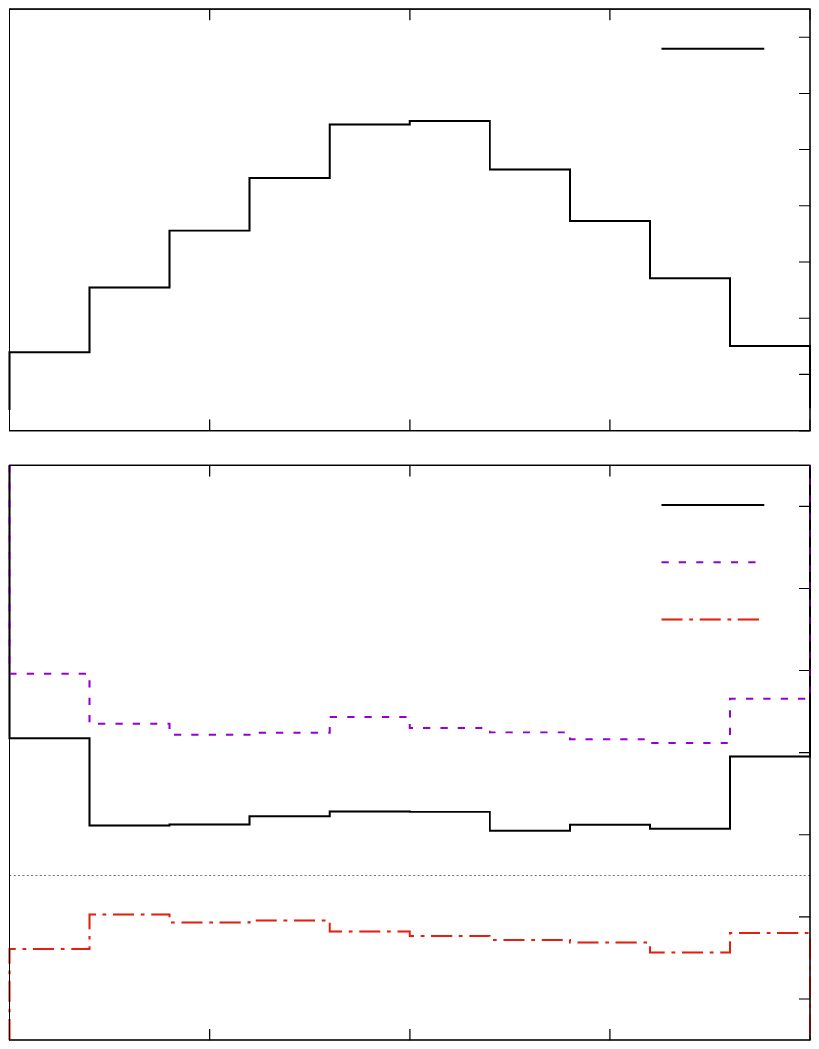}}%
    \gplfronttext
  \end{picture}%
\endgroup

%% file: sections/04_conclusion.tex
\section{Conclusion and outlook}\label{conclusion}

In this article, we
presented the calculation of mixed QCD$\otimes$QED corrections to the production of an on-shell $Z$ boson at the LHC.
We  made  use of the nested soft-collinear subtraction
scheme developed for NNLO QCD computations  at a fully-differential level, 
and extended it to cover the mixed
QCD$\otimes$QED corrections. We adopted  the abelianisation procedure  introduced in Ref.~\cite{deFlorian:2018wcj}
and included partonic channels  with photons in the initial state. Since we considered production of an on-shell
$Z$ boson, interactions between initial state partons  and decay products of the $Z$ boson can be neglected. 

As an illustration of the fully-exclusive nature of our computation, 
we calculated  the  mixed QCD$\otimes$QED corrections  to  a number of observables such as 
the transverse momentum distributions of  di-leptons  and of the  leading and subleading  leptons,
as well as the rapidity of the dilepton system and one of the Collins-Soper angles $\theta^*$. Initial-initial 
QCD$\otimes$QED corrections  typically change these distributions at below a  permille  level
whereas initial-final  ones change them by a few permille. 
This is a factor hundred (ten) smaller than the effects of NNLO QCD corrections, respectively. 

Finally, we note that the production of the  on-shell $Z$ boson is a relatively
simple case. In the future, it  may be interesting to  compute 
the mixed QCD$\otimes$QED
corrections to the  generic off-shell  Drell-Yan process and to the production
of an on-shell $W$ boson. Thanks to recent developments, both of these
computations are now feasible. The case of $W$ production will require an extension of the nested soft-collinear 
subtraction scheme to the case
of a charged resonance. This is an interesting  problem that we plan to address in the future. 

\section*{Acknowledgments}

We are grateful to A.~Behring,
F.~Caola, D.~de Florian, P.~F.~Monni, and G. Salam for useful conversations. The research of K.M. is supported by BMBF grant 05H18VKCC1 and by the DFG Collaborative Research Center TRR~257 ``Particle Physics Phenomenology after the Higgs Discovery''. M.D. and M.J. are supported
by the Deutsche Forschungsgemeinschaft (DFG, German Research Foundation) under grant 396021762 - TRR 257.